\documentclass[aps,prb,twocolumn,showpacs,amsfonts]{revtex4}

\usepackage{epsfig}
\usepackage{amsmath}
\usepackage{amssymb}
\usepackage{bm}

\newcommand{\beq}{\begin{equation}}
\newcommand{\eeq}{\end{equation}}
\newcommand{\beqarray}{\begin{eqnarray}}
\newcommand{\eeqarray}{\end{eqnarray}}

\newcommand{\Hc}{\ensuremath{\mbox{H.c.}}} % Hermitian conjugate
\newcommand{\bsig}{\mbox{\boldmath$\sigma$}} % robust bold sigma

\newcommand{\eq}[1]{Eq.~(\ref{#1})} % Eq. label
\newcommand{\fig}[1]{Fig.~\ref{#1}} % Fig. label
\newcommand{\Sec}[1]{Sec.~\ref{#1}} % Sec. label
\newcommand{\Ref}[1]{Ref.~\onlinecite{#1}} % Ref. label
\begin{document}

\allowdisplaybreaks

\title{Spin excitations in the excitonic spin-density-wave state of the
  iron pnictides}
\author{P. M. R. Brydon}
\email{brydon@theory.phy.tu-dresden.de}
\author{C. Timm}
\email{carsten.timm@tu-dresden.de}
\affiliation{Institut f\"{u}r Theoretische Physik, Technische Universit\"{a}t
  Dresden, 01062 Dresden, Germany }

\date{September 8, 2009}

\begin{abstract}
Motivated by the iron pnictides, we examine the spin excitations in an
itinerant antiferromagnet where a
spin-density wave (SDW) originates from an excitonic instability of nested
electron-like and hole-like Fermi pockets. Using the random phase
approximation (RPA),
we derive the Dyson equation for the transverse susceptibility in the
excitonic SDW state. The Dyson equation is solved for two different two-band
models, describing an antiferromagnetic insulator and metal, respectively. We
determine the collective spin-wave dispersions and also consider the
single-particle continua. The results for the excitonic models are compared
with each other and also contrasted with the well-known SDW state of the Hubbard
model. Despite the qualitatively different SDW states in the
two excitonic models, their magnetic response shows many similarities.
We conclude with a discussion of the relevance of the excitonic SDW
scenario to the iron pnictides.
\end{abstract}


\maketitle

\section{Introduction}

% pnictides

The recent discovery of superconductivity in iron pnictides has sparked a 
tremendous research effort.\cite{Kamihara2008,Rotter2008} The remarkably 
high superconducting transition temperature $T_c$ of some of these
compounds,\cite{Ren2008} their layered quasi-two-dimensional
structure,\cite{Lynn2009} the proximity
of superconductivity and antiferromagnetism in their phase
diagrams,\cite{delaCruz2008,Luetkens2009,1111coupling,122coupling} and
the likely unconventional superconducting pairing
state\cite{Mu2008,Shan2008,Si2008,Kuroki2008,Korshunov2008} are
reminiscent of the cuprates.\cite{Lee2006} It is a tantalizing
prospect that the iron pnictides can shed new light onto the problem of
unconventional high-$T_c$ superconductivity in general.

For this it is essential to assess the differences between the cuprates and
the iron pnictides. For example, the pnictides have a much more complicated
Fermi surface.\cite{Singh2008} The antiferromagnetic states in the
two families are also qualitatively different. In the cuprates,
superconductivity appears by doping an insulating antiferromagnetic parent
compound. The pnictide parent compounds \textit{R}FeAsO (\textit{R} is a
rare-earth ion) and \textit{A}Fe$_{2}$As$_{2}$ (\textit{A} is an
alkaline-earth ion) are also antiferromagnets, but there is compelling
evidence that they display a metallic SDW state: the value of the magnetic
moment at the Fe sites is small,\cite{1111coupling,122coupling,McGuire2008}
the compounds display metallic transport properties below the N\'{e}el
temperature $T_N$,\cite{McGuire2008,Liu2008,Dong2008} and ARPES and quantum
oscillation experiments find a reconstructed Fermi surface below
$T_N$.\cite{magneto,Hsieh2008}

%and it is likely that several bands participate in the
%pairing.\cite{Khasanov2009,Daghero2009,Evtushinsky2009}

The electron-phonon interaction in the pnictides is much too weak to account
for the high $T_c$ values.\cite{Boeri} Instead, the most likely candidate
for the ``glue'' binding the electrons into Cooper pairs are spin
fluctuations,\cite{Si2008,Kuroki2008,Korshunov2008} which are enhanced by the
proximity to the SDW state. A
proper understanding of the SDW phase is therefore likely the key to the
physics of the pnictides. Intriguingly,
\textit{ab-initio} calculations suggest that the nesting of electron and hole
Fermi pockets is responsible for the SDW,\cite{Singh2008}
indicating that, like the superconductivity, the antiferromagnetism of these
compounds has a multiband character. The best known material where a SDW arises
from such a nesting property is
chromium\cite{Fedders1966,Liu1970,Rice1970,Kulikov1984,FawcettCr}
and this mechanism has also been implicated for manganese
alloys.\cite{FishmanLiuMn}

%considering the multiband nature of the superconducting state,

%\cite{FishmanLiuCr}

The SDW in these compounds belongs to a broader class of density-wave
states. Consider a material with electron-like and hole-like Fermi
pockets separated by a nesting vector ${\bf Q}$ in the presence of
interband Coulomb repulsion. Performing a particle-hole transformation on one of
the bands, we obtain an attractive interaction between the particles in one
band and the holes in the other. Within a BCS-type mean-field theory, the
attractive interaction causes the condensation of
interband electron-hole pairs (excitons) with relative wave vector ${\bf
  Q}$, thereby opening a gap in the single-particle excitation
spectrum.\cite{Excitonic} Although the interband
Coulomb repulsion causes the excitonic instability,
additional interband scattering terms are required to stabilize one of several
different density-wave states, such as a SDW or a
charge-density wave (CDW).\cite{Volkov1976,Buker1981,Chubukov2008}

Several authors have discussed the SDW state of the pnictides in terms of an
excitonic instability of nested electron and hole Fermi
pockets without regard to the orbital origin of these
bands.\cite{Han2008,Mizokawa2008,Chubukov2008,Vorontsov2009,Brydon2009,Cvetkovic2009} 
An alternative school of thought emphasizes the importance of the complicated
mixing of the iron $3d$ orbitals at the Fermi energy and of the various
inter-orbital interactions.\cite{Raghu2008,Ran2009,Yu2009,Lorenzana2008}
These two approaches are not contradictory, however, since the excitonic model
can be understood as an effective low-energy theory for the
orbital models.\cite{Chubukov2008,Cvetkovic2009} Furthermore, even in an
orbital model, the SDW state is still driven by the nesting of electron and
hole Fermi pockets. Indeed, at the mean-field level all these models yield
qualitatively identical conclusions.
A conceptually different picture based on the ordering of localized
moments has also been
proposed.\cite{Si2008,Yildirim2008,Uhrig2009,Krueger2009} Although it is hard
to reconcile with the observed metallic properties\cite{McGuire2008,Liu2008} and
the moderate interaction
strengths,\cite{Kroll2008,Yang2009} this picture is consistent with several
neutron-scattering experiments.\cite{Ewings2008,Zhao2009} At present, it is
difficult to discriminate between the itinerant (excitonic) and localized
scenarios, as the dynamical spin response of the
itinerant models is unknown. It is therefore desirable to determine the spin
excitations in the excitonic SDW model.

It is the purpose of this paper to examine the transverse spin susceptibility
within the excitonic SDW state of a general two-band model. We work within the
limits of weak to moderate correlation strength, using the RPA to
construct the Dyson equation for the susceptibility. In order to
understand the generic features of the spin excitations in the excitonic SDW
state, we calculate the RPA susceptibilities for the simplest model showing
this instability. We pay particular attention to the
spin waves (magnons) and damped paramagnons. In the simplest model,
however, the SDW
state is insulating. We therefore verify the robustness of our results by
applying our theory to a system where portions of the Fermi surface
remain ungapped in the SDW phase, as in the iron pnictides. We
contrast our results for the excitonic SDW state with those for
the SDW phase of the single-band Hubbard model, which is commonly used to
describe the antiferromagnetic state of the cuprates.

The structure of this paper is as follows. We commence
in~\Sec{sec:Hubbard} with a brief review of the RPA-level results for the
transverse susceptibility in the SDW state of the Hubbard
model.\cite{Schrieffer1989,Singh1990,Chubukov1992} We then proceed 
in~\Sec{sec:excitonic} with a general discussion of the excitonic SDW state in
a two-band model and present the Dyson equation for the transverse
susceptibilities. The RPA susceptibility and spin-wave dispersion is then
calculated for the insulating and metallic excitonic SDW models
in Secs.\ \ref{subsec:perf} and \ref{subsec:imperf}, respectively.
All presented results
are calculated in the limit of zero
temperature. In order to properly compare the different models, we choose
interaction strengths such that the zero-temperature SDW gap is the
same. We conclude with a comparison with experimental results in \Sec{sec:experiments}
and a summary of our work in \Sec{sec:summary}.

\section{Hubbard Model}\label{sec:Hubbard}

\begin{figure}
  \includegraphics[width=\columnwidth,clip]{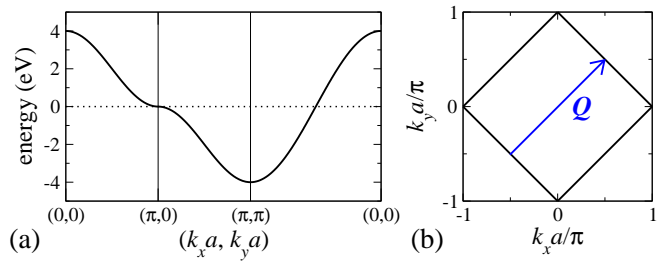}
  \caption{\label{Hub_bands}(Color online) (a) Band structure and (b)
    Fermi surface of the Hubbard model for $U=0$. In (b), the nesting vector
    ${\bf Q}=(\pi/a,\pi/a)$ is also shown.}
\end{figure}

The Hamiltonian of the Hubbard model reads
\beq
H = \sum_{{\bf k},\sigma}\epsilon_{\bf k}c^{\dagger}_{{\bf
    k},\sigma}c^{}_{{\bf k},\sigma} + \frac{U}{V}\sum_{{\bf k},{\bf k}',{\bf
    q}}c^{\dagger}_{{\bf k}+{\bf q},\uparrow}c^{}_{{\bf
    k},\uparrow}c^{\dagger}_{{\bf k}'-{\bf q},\downarrow}c^{}_{{\bf
    k}',\downarrow}, \label{eq:Hub:Ham}
\eeq
where $c^{\dagger}_{{\bf k},\sigma}$ ($c^{}_{{\bf k},\sigma}$) creates
(destroys) an electron with momentum ${\bf k}$ and spin $\sigma$.
We assume a two-dimensional (2D) nearest-neighbor tight-binding dispersion
$\epsilon_{\bf k} = -2t\,(\cos k_xa + \cos k_ya )$ where $a$ is the
lattice constant. We plot the band structure $\epsilon_\mathbf{k}$ and the
resulting Fermi surface at half filling in~\fig{Hub_bands}(a) and
(b), respectively.

At half filling and sufficiently low temperature $T$, the 
Hubbard model is unstable towards a SDW state with nesting vector ${\bf
  Q}=(\pi/a,\pi/a)$, which connects opposite sides of the Fermi
surface. We assume a SDW polarized along the $z$ axis, and
decouple the interaction term in~\eq{eq:Hub:Ham} by introducing the SDW gap
\beq
\Delta = \frac{U}{V}{\sum_{{\bf k},\sigma}}^\prime \sigma\left\langle
c^{\dagger}_{{\bf k}+{\bf Q},\sigma}c^{}_{{\bf k},\sigma} \right\rangle .
\label{eq:Hub:gap}
\eeq
The primed sum denotes summation only over the reduced, magnetic
Brillouin zone. Diagonalizing the mean-field Hamiltonian, we find two bands in
the reduced Brillouin zone with energies $E_{\pm,{\bf
    k}}=\pm\sqrt{\epsilon_{\bf k}^2+\Delta^2}$. In
the following, we will assume $t=1\,$eV and $U=0.738\,$eV, which gives a
critical temperature for the SDW state of $T_{\text{SDW}}=138\,$K and a $T=0$
gap $\Delta=21.3\,$meV.

The dynamical spin susceptibility is defined by
\beq
\chi_{ij}({\bf q},{\bf q}';i\omega_n) = \frac{1}{V}\int_{0}^{\beta}d\tau\,
\left\langle T_{\tau} 
S^{i}({\bf q},\tau)S^{j}(-{\bf
  q}',0)\right\rangle e^{i\omega_n\tau}, \label{eq:Hub:chi}
\eeq
where $T_\tau$ is the time-ordering operator and
\beq
S^{i}({\bf q},\tau) = \frac{1}{\sqrt{V}}\sum_{{\bf
k}}\sum_{s,s'}c^{\dagger}_{{\bf k}+{\bf
    q},s}(\tau)\,\frac{\sigma^{i}_{s,s'}}{2}\,c^{}_{{\bf k},s'}(\tau).
\eeq
Because of the doubling of the unit cell in the SDW state, the
susceptibility in~\eq{eq:Hub:chi} is non-zero for ${\bf q}={\bf q}'$ and
${\bf q}={\bf q}'+{\bf Q}$, the latter referred to as the umklapp
susceptibility.\cite{Schrieffer1989} Both appear in the ladder diagrams for
the transverse susceptibility, yielding the Dyson equation
\beqarray
\lefteqn{\chi_{-+}({\bf q},{\bf q}';i\omega_n) \notag} \\
& = & \delta_{{\bf q},{\bf
    q}'}\chi^{(0)}_{-+}({\bf q},{\bf q};i\omega_n) + \delta_{{\bf q}+{\bf
    Q},{\bf
    q}'}\chi^{(0)}_{-+}({\bf q},{\bf q}+{\bf Q};i\omega_n) \notag \\
&& {}+ U\chi^{(0)}_{-+}({\bf q},{\bf q};i\omega_n)\chi^{}_{-+}({\bf
  q},{\bf q}';i\omega_n) \notag \\
&& {}+ U\chi^{(0)}_{-+}({\bf q},{\bf q}+{\bf Q};i\omega_n)\chi^{}_{-+}({\bf
  q}+{\bf Q},{\bf q}';i\omega_n) ,\label{eq:Hub:Dyson}
\eeqarray
where the superscript $(0)$ indicates the mean-field susceptibilties.
Explicit expressions for
$\chi_{-+}({\bf q},{\bf q};i\omega_n)$ and $\chi_{-+}({\bf q},{\bf q} + {\bf
  Q};i\omega_n)$ can be found in~\Ref{Chubukov1992}.

\begin{figure*}
  \includegraphics[width=1.4\columnwidth]{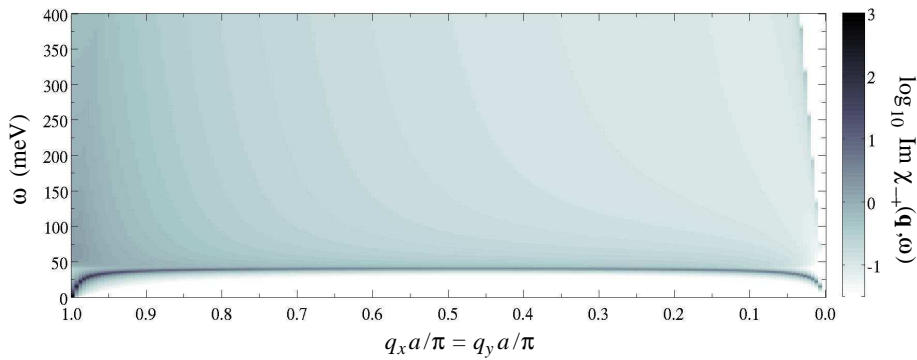}
  \caption{\label{Hubtchi3_tot}(Color online) Imaginary part of the
  transverse
  susceptibility in the Hubbard model for ${\bf q}=(q_x,q_y=q_x)$. The
  spin-wave dispersion is visible as the dark line running across the
  figure at $\omega<2\Delta=42\,$meV. Note the logarithmic color scale.}
\end{figure*}

We plot the imaginary part of $\chi_{-+}({\bf q},\omega)=\chi_{-+}({\bf
  q},{\bf q};\omega)$ along the line ${\bf q}=(q_x,q_y=q_x)$
in~\fig{Hubtchi3_tot}. The 
calculation of the mean-field susceptibilities in the Dyson
equation (\ref{eq:Hub:Dyson}) was performed over
a 10000$\times$10000 ${\bf k}$-point mesh. In the analytic continuation
$i\omega_n\rightarrow\omega+i\delta$ we
assume a finite width $\delta=1\,$meV. Smaller values of
$\delta$ and finer ${\bf k}$-point meshes do not produce
qualitative or significant quantitative changes in our results.

%In order to properly evaluate these susceptibilities,

$\mbox{Im}\,\chi_{-+}({\bf q},\omega)$ in \fig{Hubtchi3_tot}
displays very different behavior for energies
$\omega<2\Delta=42.6\,$meV and $\omega>2\Delta$. In the former region, the
dispersion of the collective spin waves is clearly
visible as the sharp dark line. The finite width of this line is
a consequence of the broadening $\delta$.
The dispersion is almost flat for
$0.1\,\pi/a\lesssim{q_x=q_y}\lesssim0.9\,\pi/a$, where it lies very close to
$\omega=2\Delta$. The distribution of spectral weight
for the spin wave is
asymmetric, with much greater weight close to ${\bf q}={\bf Q}$ than at
${\bf q}=0$, reflecting the suppression of long-wavelength
spin excitations in the SDW state.\cite{Anderson1952}

For $\omega>2\Delta$, we find a continuum of
excitations. It starts abruptly at $\omega=2\Delta$, corresponding to the
minimum energy for a single-particle excitation across the SDW gap. This
minimum is the same at all ${\bf k}$ points lying on the Fermi surface shown
in~\fig{Hub_bands}(b). By inspection, we see that for
every value of ${\bf q}$, there exist points ${\bf k}$ and ${\bf k}+{\bf q}$
lying on the Fermi surface so that the minimum energy
required for any excitation is $\omega=2\Delta$. We also see that 
$\mbox{Im}\,\chi_{-+}({\bf q},\omega)$ tends to decrease with increasing
$\omega$. This can be understood in terms of the density of states (DOS) in
the non-interacting model: the DOS has a van Hove
singularity at the Fermi energy and decreases monotonically as one moves to
higher or lower energies.
For an occupied state with energy $\omega_{o}$ below the Fermi energy, the
density of unoccupied states with energy $\omega_{u}$ above the Fermi energy
therefore decreases with increasing $\omega=\omega_{u}-\omega_{o}$, and hence
the ``density of
excitations'' contributing to the transverse susceptibility also decreases
with increasing $\omega$.

%, which for this model is also the edge of the reduced Brillouin zone in the
%SDW state

\begin{figure}
  \includegraphics[width=0.9\columnwidth,clip]{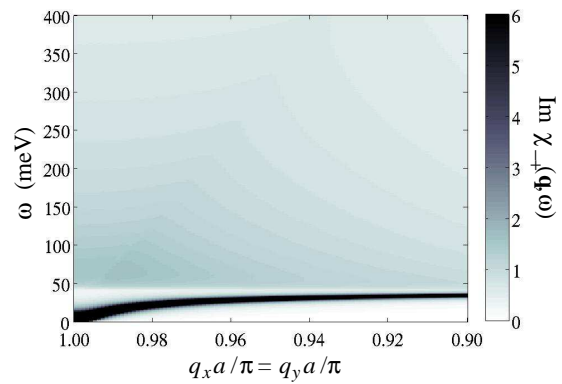}
  \caption{\label{Hubtchi3_1_lin}(Color online) Imaginary part of the
  transverse spin susceptibility in the Hubbard model for ${\bf
  q}=(q_x,q_y=q_x)$ close
  to ${\bf Q}$. Note the linear color scale.}
\end{figure}

Close to ${\bf q}=0$, the continuum is bounded from above by the line
$\omega={\bf
  v}_{F}\cdot{\bf q}$ where ${\bf v}_{F}$ is the Fermi velocity along ${\bf
  k}=(k_x,k_y=k_x)$. The peak in $\mbox{Im}\,\chi_{-+}({\bf q},\omega)$ at
this edge of the continuum is due to single-particle
excitations across the Fermi energy in the same branch of the band structure.
A rather weak dispersing feature also appears within the continuum near ${\bf
  q}={\bf Q}$, 
as shown in more detail in~\fig{Hubtchi3_1_lin}. This paramagnon originates from
single-particle excitations into the back-folded band. Like the
feature at small ${\bf q}$, the paramagnon disperses with the Fermi
velocity. The paramagnon and spin-wave dispersions curve away from
one another in what appears to be an avoided crossing.

%the importance of particle-hole-symmetric excitations for these
%$({\bf q},\omega)$ values.

%particle-hole-symmetric excitations into the back-folded band

\begin{figure}
  \includegraphics[width=\columnwidth,clip]{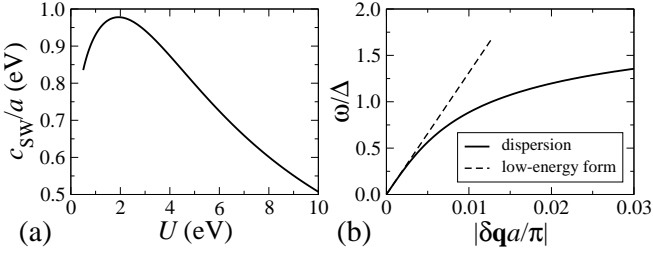}
  \caption{\label{Hubsw}(a) Spin-wave velocity $c_{\text{SW}}$ in the
    Hubbard model as a function of $U$ for $t=1\,$eV. (b) Comparison of the
    spin-wave
    dispersion and the low-energy linear form as a function of
    $\delta{\bf{q}}={\bf{Q}}-{\bf{q}}$ for $U=0.738\,$eV.}
\end{figure}

Solving~\eq{eq:Hub:Dyson} for $\chi_{-+}({\bf
  q},{\bf q};i\omega_n)$ requires the inversion of a $2\times2$
matrix. The determinant ${\cal{D}}({\bf q},i\omega_n)$ of this matrix is 
\beqarray
{\cal D}({\bf q},i\omega_n)
& = & \left[1-U\chi^{(0)}_{-+}({\bf q},{\bf
    q};i\omega_n)\right]\notag \\
&& \quad{}\times \left[1-U\chi^{(0)}_{-+}({\bf q}+{\bf Q},{\bf q}+{\bf
    Q};i\omega_n)\right] \notag \\
&& {}- \left[U\chi^{(0)}_{-+}({\bf q},{\bf q}+{\bf Q};i\omega_n)\right]^2 .
\eeqarray
Making the analytic continuation $i\omega_{n}\rightarrow\omega+i0^{+}$, the
solution of
$\mbox{Re}\,{\cal D}({\bf q},\omega)=0$ yields the spin-wave dispersion. At low
energies,
it has a linear dependence upon $\delta{\bf q} = {\bf Q} - {\bf q}$, i.e.,
$\omega=c_{\text{SW}}\,|\delta{\bf q}|$, where $c_{\text{SW}}$ is the
spin-wave
velocity. An expression for $c_\text{SW}$ is obtained by expanding
${\cal D}({\bf q},\omega)$ about ${\bf q}={\bf Q}$ and
$\omega=0$.\cite{Schrieffer1989,Singh1990,Chubukov1992} In agreement
with~\Ref{Singh1990}, we find
\beq
c_{\text{SW}} = \sqrt{\frac{-4(1/U-\Delta^2x)t^2\gamma}{x/U}}, \label{eq:Hub:csw}
\eeq
where
\beqarray
x &=& \frac{1}{V}{\sum_{\bf k}}^\prime\, \frac{1}{E_{\bf k}^3} ,\\ 
\gamma
&=&\frac{1}{V}{\sum_{\bf k}}^\prime\, \left\{\frac{1}{E^3_{\bf
    k}}\left(\cos^{2}k_xa + \cos{k_xa}\cos{k_ya}\right)\right. \notag \\ 
&& \left.+ \frac{\epsilon_{\bf k}^2-2\Delta^2}{E_{\bf
    k}^5}\sin^{2}k_{x}a\right\} . \label{eq:Hub:xandgamma}
\eeqarray
The spin-wave velocity is plotted as a function of $U$ in~\fig{Hubsw}(a), while
we compare
the low-energy linearized form of the spin-wave dispersion to the
numerically-determined result in~\fig{Hubsw}(b). As can be seen, the
linearized result holds only for small energies $\omega\lesssim 0.5\,\Delta$.
%Note that
%our expression for $\gamma$ in~\eq{eq:Hub:xandgamma} is slightly different
%from that given in~\Ref{Chubukov1992}.

\section{Excitonic Model}\label{sec:excitonic}

%and then proceed to construct the Dyson equation for the 
%transverse susceptibility. We follow this general discussion with results for
%two models of the band structure, specifically a system where the
%opening of the SDW gap
%completely removes the Fermi surface (\Sec{subsec:perf}), and a
%system where the system remains metallic in the SDW phase
%(\Sec{subsec:imperf}).

In this section we discuss the excitonic SDW in a general two-band model with
Fermi-surface nesting. We begin by outlining the known results for the
mean-field SDW
state.\cite{Excitonic,Rice1970,Volkov1976,Buker1981,Kulikov1984}
We write the Hamiltonian as
\beq
H = H_{0} + H_{I} , \label{eq:EI:Ham}
\eeq
where the non-interacting system is described by
\beq
H_{0} = \sum_{\bf k}\sum_{\sigma}\left[(\epsilon^{c}_{\bf
  k}-\mu)c^{\dagger}_{{\bf k}\sigma}c^{}_{{\bf k}\sigma} + (\epsilon^{f}_{\bf
  k}-\mu)f^{\dagger}_{{\bf k}\sigma}f^{}_{{\bf k}\sigma}\right]
\eeq
and $c^{\dagger}_{{\bf k},\sigma}$ ($f^{\dagger}_{{\bf k},\sigma}$) 
creates an electron with spin $\sigma$ and momentum ${\bf k}$ in the
electron-like $c$ band (hole-like $f$ band). The second term in~\eq{eq:EI:Ham}
describes the interactions in the model
system. Following~Ref.s~\onlinecite{Chubukov2008}
and~\onlinecite{Cvetkovic2009},  
we take this to consist of five on-site terms $H_{I} = H_{cc} + H_{f\!f} +
H_{cf} + H_\text{ITa} + H_\text{ITb}$ that arise naturally in the low-energy
effective theory of a multi-orbital model. These correspond to intraband Coulomb
repulsion,
\beqarray
H_{cc} & = & \frac{g_{cc}}{V}\sum_{{\bf k},{\bf k}',{\bf
    q}}c^{\dagger}_{{\bf k}+{\bf q},\uparrow}c^{}_{{\bf
    k},\uparrow}c^{\dagger}_{{\bf k}'-{\bf q},\downarrow}c^{}_{{\bf
    k}',\downarrow} ,\\
H_{f\!f} &=& \frac{g_{f\!f}}{V}\sum_{{\bf k},{\bf k}',{\bf
    q}}f^{\dagger}_{{\bf k}+{\bf q},\uparrow}f^{}_{{\bf
    k},\uparrow}f^{\dagger}_{{\bf k}'-{\bf q},\downarrow}f^{}_{{\bf
    k}',\downarrow} ,
\eeqarray
interband Coulomb repulsion,
\beq
H_{cf} = \frac{g_{cf}}{V}\sum_{{\bf k},{\bf k}',{\bf
    q}}\sum_{\sigma,\sigma'}c^{\dagger}_{{\bf k}+{\bf q},\sigma}c^{}_{{\bf
    k},\sigma}f^{\dagger}_{{\bf k}'-{\bf q},\sigma'}f^{}_{{\bf
    k}',\sigma'}, \label{eq:EI:InterBCoulomb}
\eeq
and two distinct types of correlated interband transitions,
\begin{align}
H_\text{ITa} &=\,  \frac{g_\text{2a}}{V}\sum_{{\bf k},{\bf k}',{\bf
    q}}\left(c^{\dagger}_{{\bf k}+{\bf
    q},\uparrow}c^{\dagger}_{{\bf k}'-{\bf q},\downarrow}f^{}_{{\bf
    k}',\downarrow}f^{}_{{\bf k},\uparrow} + \Hc\right), \\
H_\text{ITb} & =\, \frac{g_\text{2b}}{V}\sum_{{\bf k},{\bf k}',{\bf
    q}}\sum_{\sigma,\sigma'}c^{\dagger}_{{\bf k}+{\bf
    q},\sigma}f^{\dagger}_{{\bf k}'-{\bf q},\sigma'}c^{}_{{\bf
    k}',\sigma'}f^{}_{{\bf k},\sigma}.
\end{align}
The interband interaction terms are responsible for a density-wave instability
when the electron and hole Fermi pockets are sufficiently close to nesting. A
number of different density-wave states are possible:\cite{Buker1981} a
CDW with effective coupling constant
$g_{\text{CDW}}=g_{cf}-g_\text{2a}-2g_\text{2b}$, a SDW with coupling
$g_{\text{SDW}}=g_{cf}+g_\text{2a}$,
a charge-current-density wave (CCDW) with
$g_{\text{CCDW}}=g_{cf}+g_\text{2a}-2g_\text{2b}$,
and a spin-current-density wave (SCDW) with $g_{\text{SCDW}}=g_{cf}-g_\text{2a}$.
In order to model the iron pnictides,
we henceforth assume that the SDW state has the largest coupling constant.

%Although the excitonic instability has been implicated in
%several materials showing a CDW or
%SDW,\cite{Kulikov1984,FawcettCr,Ta2NiSe5,TiSe2} we are not aware of any
%examples of the excitonic current density wave states.

In the presence of a SDW polarized along the $z$-axis and with nesting vector
${\bf Q}$, the effective mean-field Hamiltonian is written as
\beq
H_\text{MF} = H_{0} + \sum_{\bf
  k}\sum_\sigma\sigma\Delta\left(c^{\dagger}_{{\bf k},\sigma}f^{}_{{\bf
    k}+{\bf Q},\sigma} + \Hc\right) ,
\eeq
where the excitonic gap
\beq
\Delta = \frac{g_{\text{SDW}}}{2V}\sum_{\bf k}\sum_{\sigma}\sigma\left\langle
c^{\dagger}_{{\bf k},\sigma}f^{}_{{\bf k}+{\bf Q},\sigma} \right\rangle
\eeq
is assumed to be real.
%Although $\Delta$ is the order parameter of the excitonic SDW state,
The precise relationship between $\Delta$ and the magnetization is somewhat
complicated.\cite{Volkov1976,Buker1981} To elucidate it, we define
the field operator,
\beq
\psi_{\sigma}({\bf r}) = \frac{1}{\sqrt{V}}\sum_{\bf k}\left[\varphi_{{\bf
    k},c}({\bf r})c^{}_{{\bf
    k},\sigma} + \varphi_{{\bf k},f}({\bf r})f^{}_{{\bf k},\sigma}
\right]e^{i{\bf k}\cdot{\bf r}},
\eeq
where $\varphi_{{\bf k},\alpha}({\bf r})$ is a Bloch function for the
band $\alpha$. The local magnetization ${\bf M}({\bf r})$ is then
\beqarray
{\bf M}({\bf r}) & = & -\frac{g\mu_{B}}{V}\,\sum_{s,s'}\sum_{{\bf k},{\bf
    k}'}\sum_{a,b=c,f} \varphi^{\ast}_{{\bf k},a}({\bf r})\varphi_{{\bf
    k}',b}({\bf r}) \notag \\
&& {}\times  e^{-i({\bf k} - {\bf k}')\cdot{\bf r}}
    \left\langle a^{\dagger}_{{\bf 
    k},s}\,\frac{\bsig_{s,s'}}{2}\,b^{}_{{\bf k}',s'} \right\rangle  ,
\eeqarray
where $g$ is the \textit{g}-factor and $\mu_B$ is the Bohr magneton.
Only in the limit when $\varphi_{{\bf k},\alpha}({\bf r})$ is constant do we
find $\Delta$ to be simply related to the magnetization,
\beq
{\bf M}({\bf r}) =
-\frac{2g\mu_{B}\Delta}{g_{\text{SDW}}}\,
  (\cos {\bf Q}\cdot{\bf r})\,{\bf e}_z .
\eeq
For simplicity, we follow
Refs.~\onlinecite{Liu1970,Hasegawa1978,FishmanLiuMn,Korshunov2008} in assuming
constant Bloch functions.

%\cite{FishmannLiuCr}

In calculating the susceptibilities, we make use of the
single-particle
Green's functions of the mean-field SDW state. The two normal
(diagonal in band indices) Green's functions are defined by
\beqarray
G^{cc}_{{\bf k},\sigma}(i\omega_n) & = & -\int^{\beta}_{0} d\tau \left\langle
T_{\tau}c^{}_{{\bf k},\sigma}(\tau)c^{\dagger}_{{\bf
    k},\sigma}(0)\right\rangle e^{i\omega_{n}\tau} \notag \\
& = & \frac{i\omega_n - \epsilon^{f}_{{\bf k}+{\bf Q}}}{(i\omega_n - E_{+,{\bf
      k}+{\bf Q}})(i\omega_n - E_{-,{\bf k}+{\bf Q}})}, \label{eq:EI:Gcc} \\
G^{f\!f}_{{\bf k},\sigma}(i\omega_n) & = & -\int^{\beta}_{0} d\tau \left\langle
T_{\tau}f^{}_{{\bf k},\sigma}(\tau)f^{\dagger}_{{\bf
    k},\sigma}(0)\right\rangle e^{i\omega_{n}\tau} \notag \\
& = & \frac{i\omega_n - \epsilon^{c}_{{\bf k}+{\bf Q}}}{(i\omega_n - E_{+,{\bf
      k}})(i\omega_n - E_{-,{\bf k}})}, \label{eq:EI:Gff} 
\eeqarray
while the anomalous (band-mixing) Green's functions are
\beqarray
G^{fc}_{{\bf k},\sigma}(i\omega_n) & = & -\int^{\beta}_{0} d\tau \left\langle
T_{\tau}f^{}_{{\bf k},\sigma}(\tau)c^{\dagger}_{{\bf
    k}+{\bf Q},\sigma}(0)\right\rangle e^{i\omega_{n}\tau} \notag \\
& = & \frac{\sigma\Delta}{(i\omega_n - E_{+,{\bf
      k}})(i\omega_n - E_{-,{\bf k}})},  \label{eq:EI:Gfc} \\
G^{cf}_{{\bf k},\sigma}(i\omega_n) & = & -\int^{\beta}_{0} d\tau \left\langle
T_{\tau}c^{}_{{\bf k},\sigma}(\tau)f^{\dagger}_{{\bf
    k}+{\bf Q},\sigma}(0)\right\rangle e^{i\omega_{n}\tau} \notag \\
& = & G^{fc}_{{\bf k}+{\bf Q},\sigma}(i\omega_n).
 \label{eq:EI:Gcf} 
\eeqarray
The functions $E_{\pm,{\bf k}}$ are the dispersion relations for the
reconstructed bands,
\beq
E_{\pm,{\bf k}} = \frac{1}{2}\left[\epsilon^{c}_{{\bf k}+{\bf Q}} +
\epsilon^{f}_{{\bf
      k}}
  \pm \sqrt{\left(\epsilon^{c}_{{\bf k} + {\bf Q}} - \epsilon^{f}_{{\bf
k}}\right)^2 +
    4\Delta^2}\right]. \label{eq:EI:rebuilt}
\eeq
For energies much larger than $\Delta$ we have $E_{+,{\bf
    k}}\approx\epsilon^{c}_{{\bf k}+{\bf Q}}$
and $E_{-,{\bf k}}\approx\epsilon^{f}_{{\bf k}}$.

\begin{figure*}
  \includegraphics[width=2\columnwidth]{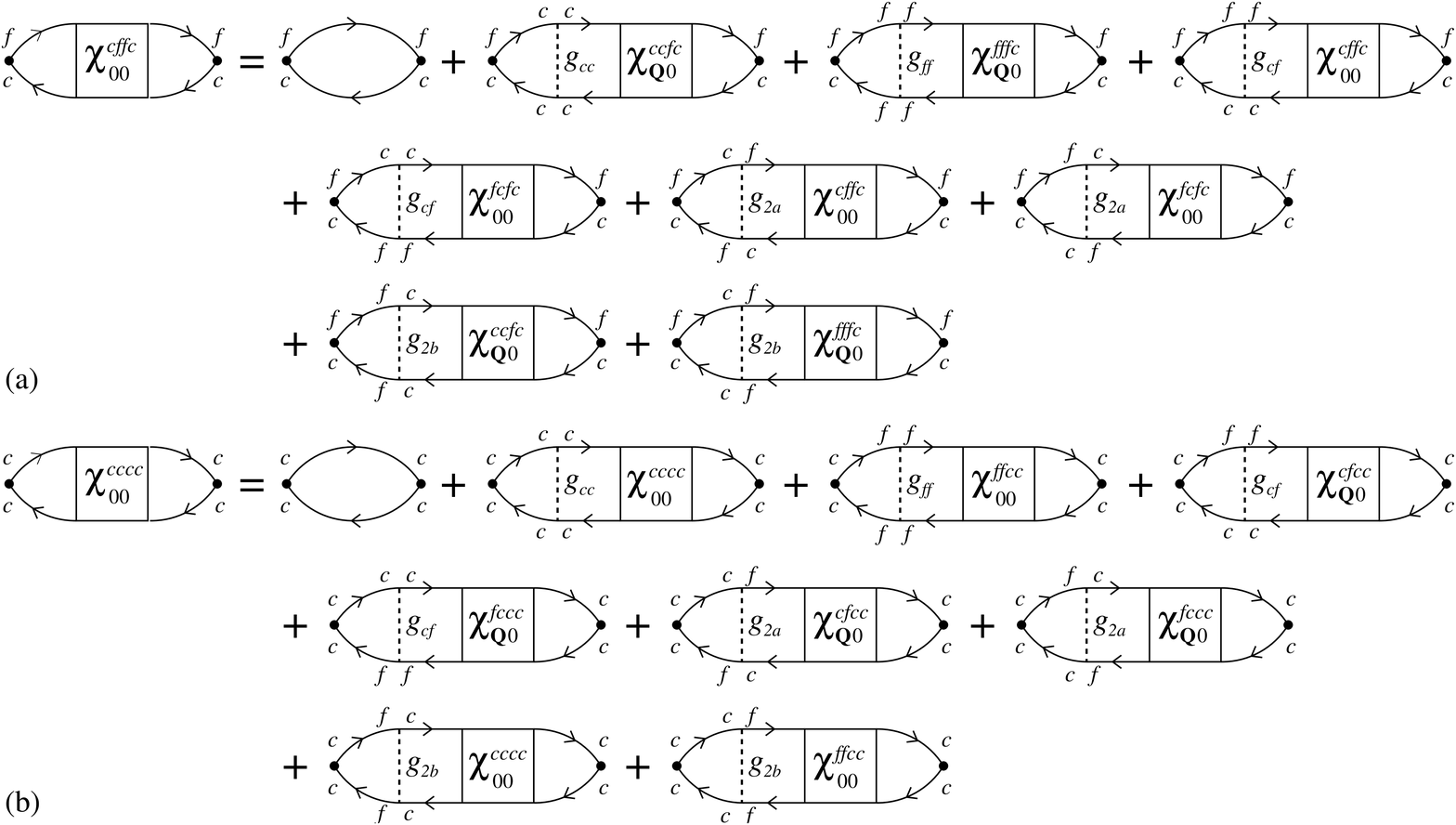}
  \caption{\label{Dyson}Diagrammatic representation of the Dyson
    equation (\ref{eq:EI:Dyson}) for (a) $\chi^{cf\!fc}_{-+,00}$ and (b)
    $\chi^{cccc}_{-+,00}$.
    The curved lines are the mean-field Green's functions in the SDW
    state. When the label $b=c,f$ follows the label $a=c,f$ in the
    direction of the arrow, the corresponding Green's function is $G^{ab}$.} 
\end{figure*}

The total spin operator is written as
\beqarray
{\bf S}({\bf r}) &=& \sum_{s,s'}\psi^{\dagger}_{s}({\bf
  r})\,\frac{\bsig_{s,s'}}{2}\,\psi^{}_{s'}({\bf r}) \notag \\
&=& \frac{1}{2V}\sum_{a,b=c,f}\sum_{{\bf k},{\bf
    q}}\sum_{s,s'}a^{\dagger}_{{\bf k}+{\bf
    q},s}\bsig_{s,s'}b^{}_{{\bf k},s'}e^{-i{\bf q}\cdot{\bf r}}
\notag \\
& = & \frac{1}{\sqrt{V}}\sum_{a,b=c,f}\sum_{\bf q}{\bf S}_{a,b}({\bf
  q})e^{-i{\bf q}\cdot{\bf r}} ,
\eeqarray
where ${\bf S}_{a,b}({\bf q})$ is a generalized spin operator. The
dynamical spin susceptibility is then defined by
\beqarray
\lefteqn{\chi_{ij}({\bf q},{\bf q}';i\omega_n) \notag }\\
&=&
\frac{1}{V}\sum_{a,b}\sum_{a',b'}\int_{0}^{\beta} \left\langle T_{\tau} 
S^{i}_{a,b}({\bf q},\tau)S^{j}_{a',b'}(-{\bf
  q}',0)\right\rangle e^{i\omega_n\tau} \notag \\
&= & \sum_{a,b}\sum_{a',b'} \chi^{aba'b'}_{ij}({\bf q},{\bf q}';i\omega_n).
\label{eq:EI:chi}
\eeqarray 
The generalized susceptibilities $\chi^{aba'b'}_{ij}({\bf q},{\bf
  q}';i\omega_n)$ are calculated using the RPA. We are only
concerned with the transverse susceptibility, which is obtained by
summing the ladder diagrams. This yields the Dyson equation
\beqarray
\lefteqn{ \chi^{aba'b'}_{-+,\,00}\notag }\\ & = & \delta_{{\bf q},{\bf
    q}'}\left(\delta_{a',b}\delta_{b',a}\chi^{abba\,(0)}_{-+,\,00}
+
\delta_{a',\overline{b}}\delta_{b',\overline{a}}\chi^{ab\overline{b}\overline{
a}\,(0)}_{-+,\,00}\right)
\notag \\
&& {}+ \delta_{{\bf q}+{\bf Q},{\bf
  q}'}\left(\delta_{a',b}\delta_{b',\overline{a}}\chi^{abb\overline{a}\,(0)}_{
-+ ,\,0{\bf
  Q}}
+
\delta_{a',\overline{b}}\delta_{b',a}\chi^{ab\overline{b}a\,(0)}_{-+,\,0{\bf
  Q}}\right) \notag \\
&& {}+
g_{cc}\left(\chi^{abcc\,(0)}_{-+,\,00}\chi^{cca'b'}_{-+,\,00}
+ \chi^{abcc\,(0)}_{-+,\,0{\bf
    Q}}\chi^{cca'b'}_{-+,\,{\bf Q}0}\right) \notag \\
&& {}+
g_{f\!f}\left(\chi^{abf\!f\,(0)}_{-+,\,00}\chi^{f\!fa'b'}_{-+,\,00}
+ \chi^{abf\!f\,(0)}_{-+,\,0{\bf
    Q}}\chi^{f\!fa'b'}_{-+,\,{\bf Q}0}\right) \notag \\
&& {}+
g_{cf}\left(\chi^{abcf\,(0)}_{-+,\,00}\chi^{fca'b'}_{-+,\,00}
+ \chi^{abcf\,(0)}_{-+,\,0{\bf
    Q}}\chi^{fca'b'}_{-+,\,{\bf Q}0} \right. \notag \\
&& \quad\left. {}+ \chi^{abfc\,(0)}_{-+,\,00}\chi^{cfa'b'}_{-+,\,00}
+ \chi^{abfc\,(0)}_{-+,\,0{\bf
    Q}}\chi^{cfa'b'}_{-+,\,{\bf Q}0}\right) \notag \\
&& {}+
g_\text{2a}\left(\chi^{abcf\,(0)}_{-+,\,00}\chi^{cfa'b'}_{-+,\,00}
+ \chi^{abcf\,(0)}_{-+,\,0{\bf
    Q}}\chi^{cfa'b'}_{-+,\,{\bf Q}0} \right. \notag \\
&& \quad\left. {}+ \chi^{abfc\,(0)}_{-+,\,00}\chi^{fca'b'}_{-+,\,00}
+ \chi^{abfc\,(0)}_{-+,\,0{\bf
    Q}}\chi^{fca'b'}_{-+,\,{\bf Q}0}\right) \notag \\
&& {}+
g_\text{2b}\left(\chi^{abcc\,(0)}_{-+,\,00}\chi^{f\!fa'b'}_{-+,\,00}
+ \chi^{abcc\,(0)}_{-+,\,0{\bf
    Q}}\chi^{f\!fa'b'}_{-+,\,{\bf Q}0} \right. \notag \\
&& \quad\left. {}+ \chi^{abf\!f\,(0)}_{-+,\,00}\chi^{cca'b'}_{-+,\,00}
+ \chi^{abf\!f\,(0)}_{-+,\,0{\bf
    Q}}\chi^{cca'b'}_{-+,\,{\bf Q}0}\right), \label{eq:EI:Dyson}
\eeqarray
where we have adopted the short-hand notation
\beqarray
\chi^{aba'b'\,(0)}_{-+,\,{\bf m}{\bf n}} & = &
\chi^{aba'b'\,(0)}_{-+}({\bf q}+{\bf m},{\bf q}+{\bf
  n};i\omega_n), \\
\chi^{aba'b'}_{-+,\,{\bf m}{\bf n}} & = &
\chi^{aba'b'}_{-+}({\bf q}+{\bf m},{\bf q}'+{\bf
  n};i\omega_n).
\eeqarray
%and all susceptibilities in~\eq{eq:EI:Dyson} are functions of
%$i\omega_{n}$.
Note that $\chi^{aba'b'\,(0)}_{-+,\,{\bf m}{\bf n}}$ does not depend on
${\bf q}'$.
We have also introduced the notation $\overline{a}=c\,(f)$
when $a=f\,(c)$.
The first two lines of~\eq{eq:EI:Dyson} give the
mean-field susceptibilties, obtained by using Wick's theorem to contract the
correlation function in~\eq{eq:EI:chi} into products of two mean-field Green's
functions. On the first line of~\eq{eq:EI:Dyson} we find the correlators
resulting from the product of two normal Green's functions,
\beq
\chi^{abba\,(0)}_{-+,\,00}
 = -\frac{1}{V}\,\sum_{\bf
  k}\frac{1}{\beta}\sum_{i\nu_{n}}G^{bb}_{{\bf
  k},\uparrow}(i\nu_{n})G^{aa}_{{\bf 
  k}+{\bf q},\downarrow}(i\nu_{n}-i\omega_{n}), \notag
\eeq
and from the product of two anomalous Green's functions,
\beq
\chi^{ab\overline{b}\overline{a}\,(0)}_{-+,\,00} = 
-\frac{1}{V}\,\sum_{\bf 
  k}\frac{1}{\beta}\sum_{i\nu_{n}}G^{b\overline{b}}_{{\bf
  k},\uparrow}(i\nu_{n})G^{\overline{a}a}_{{\bf 
  k}+{\bf q},\downarrow}(i\nu_{n}-i\omega_{n}). \notag
\eeq
On the second line we find the umklapp susceptibilties which, as in the
Hubbard model, are the product of a normal and an anomalous Green's function:
\begin{align}
\chi^{abb\overline{a}\,(0)}_{-+,\,0{\bf
    Q}} &=\, -\frac{1}{V}\,\sum_{\bf 
  k}\frac{1}{\beta}\sum_{i\nu_{n}}G^{bb}_{{\bf
  k},\uparrow}(i\nu_{n})G^{\overline{a}a}_{{\bf
  k}+{\bf q},\downarrow}(i\nu_{n}-i\omega_{n}), \notag \\
\chi^{ab\overline{b}a\,(0)}_{-+,\,0{\bf
    Q}} &=\, -\frac{1}{V}\,\sum_{\bf 
  k}\frac{1}{\beta}\sum_{i\nu_{n}}G^{b\overline{b}}_{{\bf
  k},\uparrow}(i\nu_{n})G^{aa}_{{\bf 
  k}+{\bf q},\downarrow}(i\nu_{n}-i\omega_{n}). \notag
\end{align}
The remaining lines of~\eq{eq:EI:Dyson} give the ladder sums for the various
interactions: on the third and fourth lines we have the intraband Coulomb
interactions, on
the fifth and sixth lines the interband Coulomb interaction, and on
the last four
lines the two types of correlated transitions.
In~\fig{Dyson}(a) and (b), we show a diagrammatic representation of the Dyson
equation for $\chi^{cf\!fc}_{-+,00}$ and $\chi^{cccc}_{-+,00}$, respectively.
Note that the Dyson equation is also valid in the normal state,
in which case the
$\chi^{abba\,(0)}_{-+,\,00}$ are
the only non-zero mean-field susceptibilities.

%The boxes represent the full susceptibilities.

From the structure of the Dyson equation, we see that
$\chi^{aba'b'}_{-+,{\bf mn}}$ is only non-zero for ${\bf q}={\bf q}'$ and ${\bf
  m},\,{\bf n}\in\{0,\,{\bf Q}\}$. We observe 
that~\eq{eq:EI:Dyson} may then be written as four independent sets of coupled
equations for
\begin{subequations}
\label{eq:EI:allsets}
\begin{align}
&\left\{ \chi^{cccc}_{-+,\,00}, \,\chi^{f\!fcc}_{-+,\,00},
\,\chi^{fccc}_{-+,\,{\bf
    Q}0}, \,\chi^{cfcc}_{-+,\,{\bf Q}0} \right\}, \label{eq:EI:chixxcc} \\
&\left\{ \chi^{f\!f\!f\!f}_{-+,\,00}, \,\chi^{ccf\!f}_{-+,\,00},
\,\chi^{cf\!f\!f}_{-+,\,{\bf
    Q}0}, \,\chi^{fcf\!f}_{-+,\,{\bf Q}0} \right\}, \label{eq:EI:chixxff} \\
&\left\{ \chi^{fccf}_{-+,\,00}, \,\chi^{cfcf}_{-+,\,00},
\,\chi^{cccf}_{-+,\,{\bf
    Q}0}, \,\chi^{f\!fcf}_{-+,\,{\bf Q}0} \right\}, \label{eq:EI:chixxcf} \\
&\left\{ \chi^{cf\!fc}_{-+,\,00}, \,\chi^{fcfc}_{-+,\,00},
\,\chi^{ccfc}_{-+,\,{\bf
    Q}0}, \,\chi^{f\!f\!fc}_{-+,\,{\bf Q}0} \right\}. \label{eq:EI:chixxfc}
\end{align}
\end{subequations}
Note that this includes $\chi^{aba'b'}_{-+,{\bf mn}}$ for ${\bf mn}={\bf QQ}$
and ${\bf mn}={0{\bf Q}}$ by symmetry; all other possible transverse
susceptibilities vanish. The first two
sets contain the contributions to the intraband susceptibility, which involve
spin-flip transitions within the $c$ and $f$ bands:
\beqarray
\lefteqn{\chi^{\text{intra}}_{-+}({\bf q},i\omega_n) } \notag \\& = &
\frac{1}{V}\sum_{a,b=c,f}\int^{\beta}_{0} d\tau \left\langle T_{\tau}
S^{-}_{a,a}({\bf
  q},\tau)S^{+}_{b,b}(-{\bf 
  q},0)\right\rangle e^{i\omega_{n}\tau} \notag \\
& = &
\chi^{cccc}_{-+,\,00} + \chi^{ccf\!f}_{-+,\,00} + \chi^{f\!fcc}_{-+,\,00}  +
\chi^{f\!f\!f\!f}_{-+,\,00} .
\eeqarray
The last two
sets contain the contributions to the interband susceptibility, which involve
spin-flip transitions between the $c$ and $f$ bands:
\beqarray
\lefteqn{\chi^{\text{inter}}_{-+}({\bf q},i\omega_n) } \notag \\
& = & \frac{1}{V}\sum_{a,b=c,f}\int^{\beta}_{0} d\tau\left\langle T_{\tau}
S^{-}_{a,\overline{a}}({\bf q},\tau)S^{+}_{b,\overline{b}}(-{\bf
  q},0)\right\rangle e^{i\omega_{n}\tau} \notag \\
& = & \chi^{cf\!fc}_{-+,\,00} + \chi^{cfcf}_{-+,\,00} + \chi^{fcfc}_{-+,\,00} +
\chi^{fccf}_{-+,\,00}  .
\eeqarray
We note that the Dyson equation for the interband susceptibilities has been
previously obtained in Refs.~\onlinecite{Liu1970}
and~\onlinecite{Hasegawa1978} for the case where only the interband Coulomb
interaction~\eq{eq:EI:InterBCoulomb} 
is non-zero.
From~\eq{eq:EI:chi} we see that the total transverse susceptibility is the sum
of the intraband and interband contributions,
\beq
\chi_{-+}({\bf q},i\omega_n) = \chi^{\text{intra}}_{-+}({\bf q},i\omega_n) +
\chi^{\text{inter}}_{-+}({\bf q},i\omega_n).
\eeq
Since the interband and intraband susceptibilities involve qualitatively
different types of excitations, considering these seperately offers greater
physical insight into the magnetic response than the total
susceptibility.

In the following sections we discuss the transverse susceptibility for two
different models of the band
structure. For simplicity, we restrict ourselves to the case
$g_{cc}=g_{f\!f}=g_\text{2b}=0$, as these interactions do not drive the
SDW
instability. We emphasize, however, that the
preceeding results are valid for any choice of couplings in
both the normal and SDW states. Except where stated otherwise, we furthermore
set $g_\text{2a}=0$, as at reasonable coupling strengths we find very little
change
in the transverse susceptibility upon varying $g_{cf}$ and $g_\text{2a}$ while
keeping
$g_{\text{SDW}}=g_{cf}+g_\text{2a}$ fixed. Unless explicitly mentioned,
we have used a 10000$\times$10000 ${\bf k}$-point mesh and a width
$\delta=1\,$meV to calculate the mean-field susceptibilities.

\subsection{Insulating SDW state}\label{subsec:perf}

\begin{figure}
  \includegraphics[width=\columnwidth,clip]{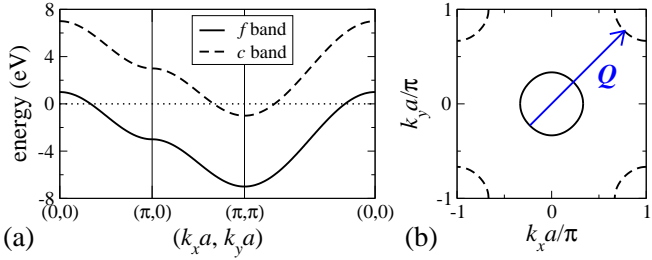}
  \caption{\label{EI1_bands}(Color online) (a) Band structure and (b)
    Fermi surface of the non-interacting insulating excitonic model. In
    (b), the nesting vector ${\bf Q}=(\pi/a,\pi/a)$ is also shown.} 
\end{figure}

We first examine an excitonic model with perfect nesting between the electron
and hole bands, i.e., $\epsilon^{c}_{{\bf k}}=-\epsilon^{f}_{{\bf
    k}+{\bf Q}}$ for all ${\bf k}$. Although hardly realistic, at the
mean-field level it exactly maps onto the BCS model after particle-hole
transformation.\cite{Excitonic} It is
therefore useful for obtaining physically transparent
results and is frequently encountered in the
literature.\cite{Volkov1976,Excitonic,Chubukov2008,Vorontsov2009,Cvetkovic2009}
We assume the 2D band structure
\begin{subequations}
\label{eq:EI:perfnest_disp}
\beqarray
\epsilon^{c}_{\bf k} & = & 2t\left(\cos{k_xa} + \cos{k_ya}\right) +
\epsilon_{0}, \\
\epsilon^{f}_{\bf k} & = & 2t\left(\cos{k_xa} + \cos{k_ya}\right) -
\epsilon_{0},
\eeqarray
\end{subequations}
where we set $t=1\,$eV and $\epsilon_{0}=3\,$eV. The band structure and Fermi
surface at half-filling are shown in \fig{EI1_bands}(a) and (b),
respectively. Below we will take $g_{\text{SDW}}=1.8\,$eV, for which the
mean-field
equations yield a SDW with nesting vector ${\bf Q}=(\pi/a,\pi/a)$, critical
temperature $T_{\text{SDW}}=138\,$K, and $T=0$ gap $\Delta=21.3\,$meV. The
system is insulating at $T=0$, with the SDW gap completely removing the Fermi
surface.

\begin{figure*}
  \includegraphics[width=1.4\columnwidth]{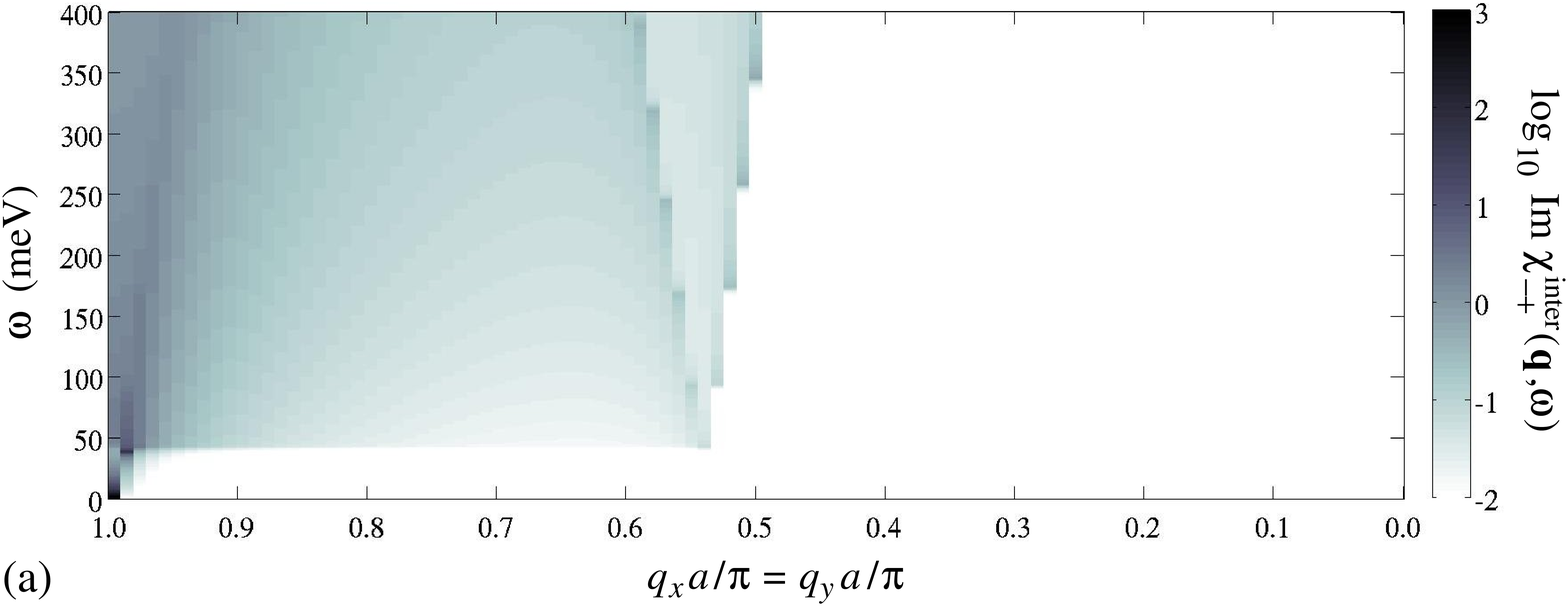}
  \includegraphics[width=1.4\columnwidth]{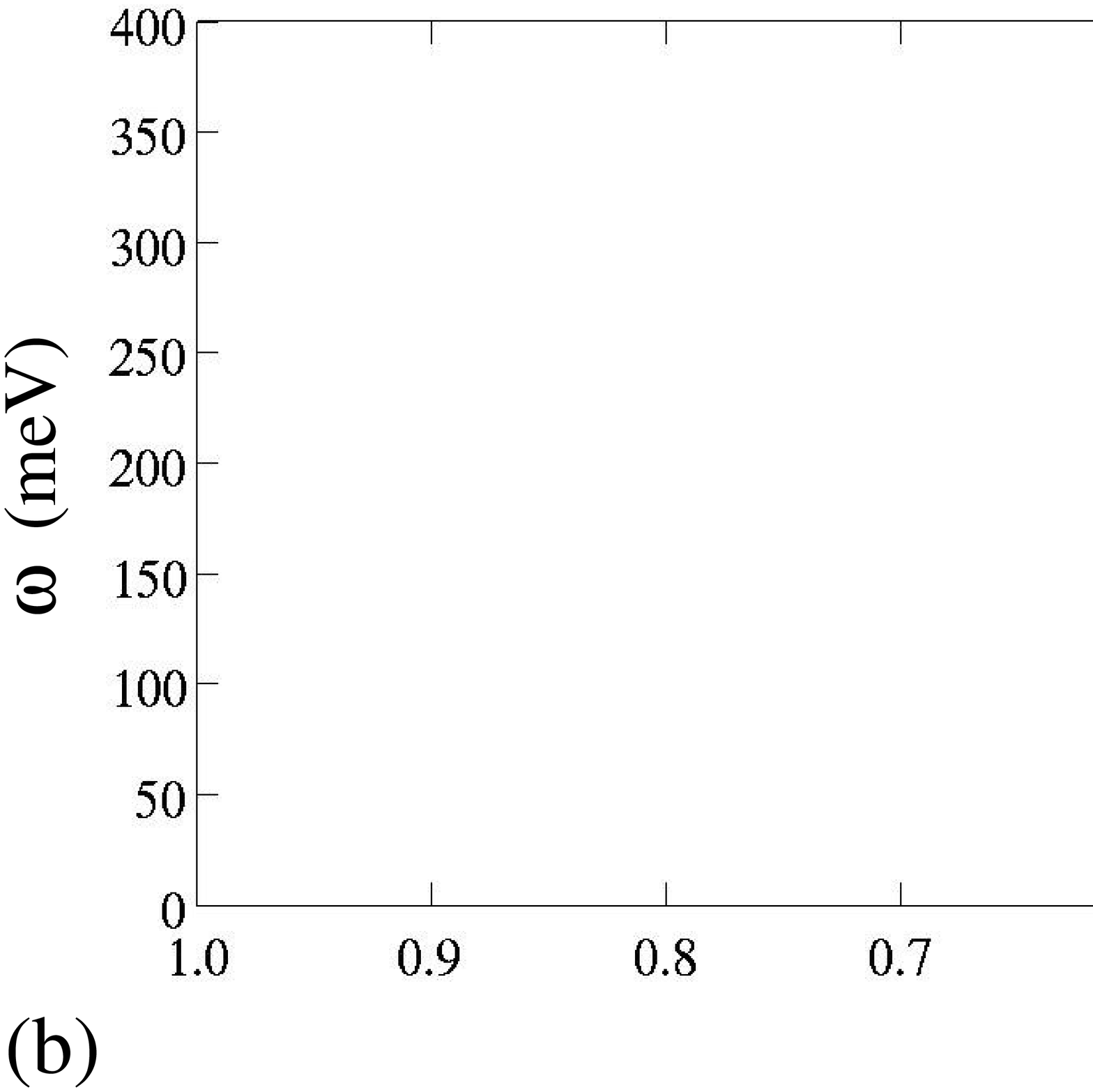}
  \includegraphics[width=1.4\columnwidth]{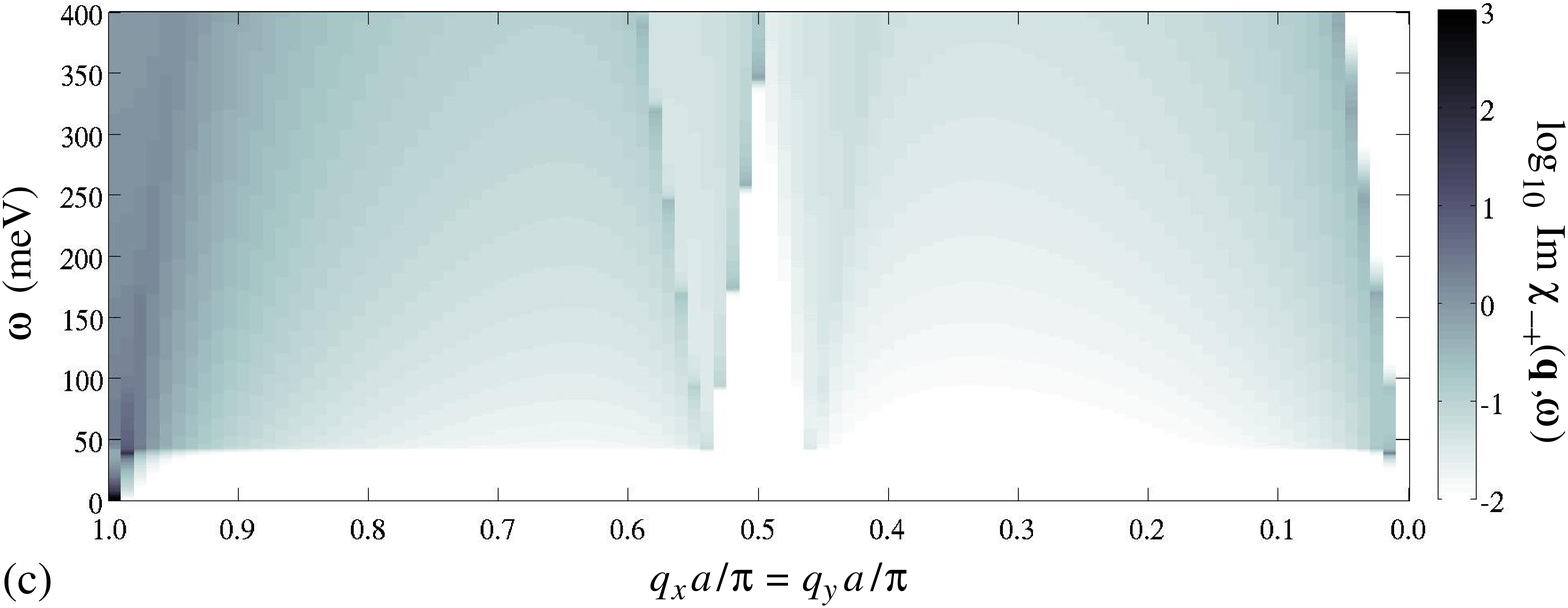}
  \caption{\label{perfnest_chi}(Color online) Imaginary part of (a) the
    interband, (b) the intraband, and (c) the total transverse
  susceptibility in the insulating excitonic model for ${\bf
    q}=(q_x,q_y=q_x)$. The spin-wave dispersion is visible as the dark
  line at $\omega<2\Delta$ close to ${\bf Q}$ in (a) and (c). Note
  the logarithmic color scale.}
\end{figure*}

\begin{figure}
(a)\includegraphics[width=0.9\columnwidth]{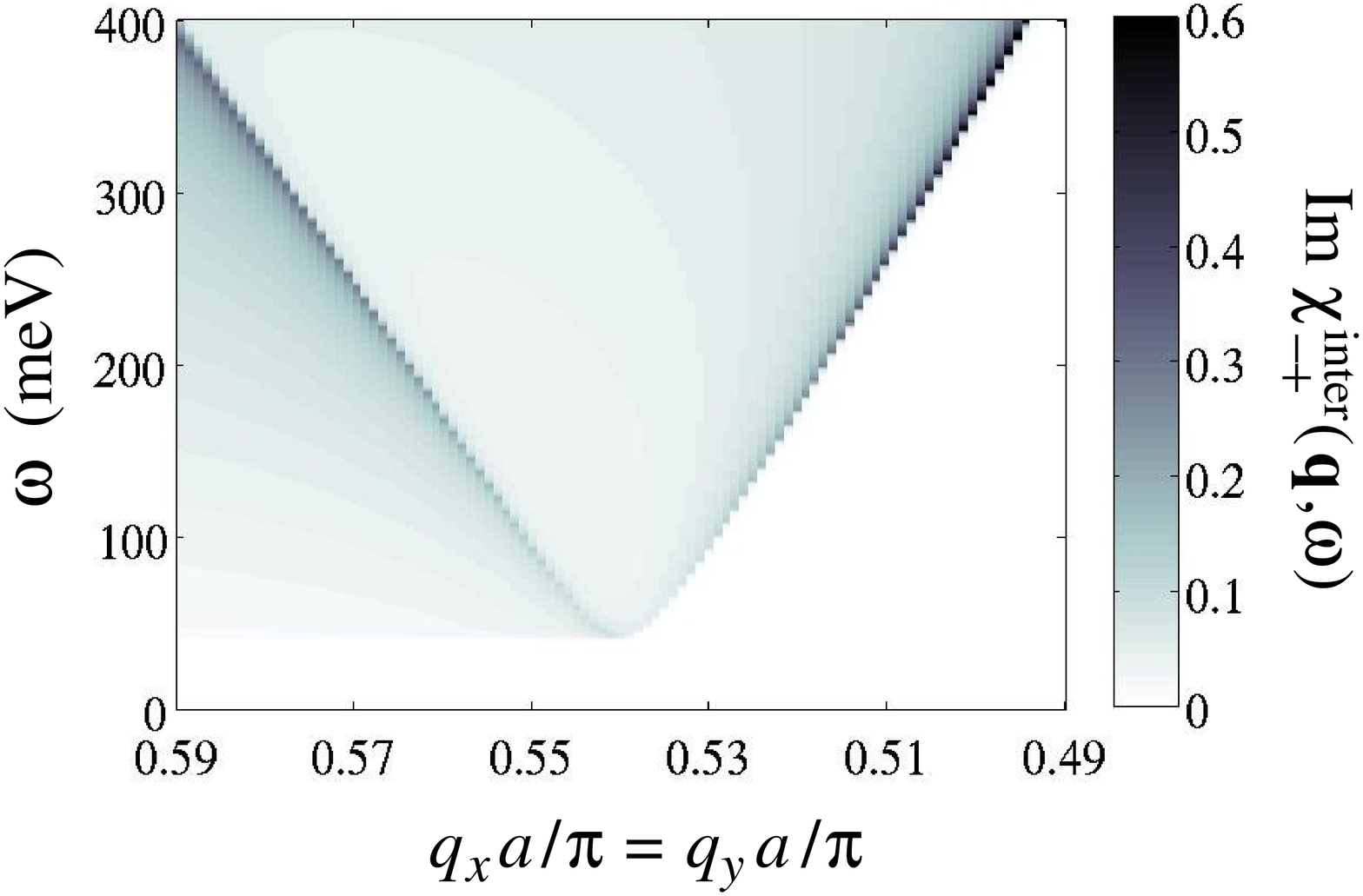}\\
(b)\includegraphics[width=0.9\columnwidth,clip]{fig8b.eps}
  \caption{\label{interbchi13}(Color online) (a) Imaginary part of the
    interband transverse susceptibility in the insulating excitonic
    model for ${\bf q}=(q_x,q_y=q_x)$ close to $0.54\,{\bf Q}$. Note the linear
    color scale. (b) Band structure along the Brillouin zone diagonal, showing
    the nesting responsible for the dispersing feature in (a).}
\end{figure}

We plot the imaginary parts of the interband, intraband, and total transverse
susceptibilties for ${\bf q}=(q_x,q_y=q_x)$
in~\fig{perfnest_chi}(a), (b), and (c), respectively. We consider first the
interband contribution. For ${\bf q}$ sufficiently close to ${\bf Q}$, we find
a continuum of single-particle excitations.
In contrast to the
results for the Hubbard model (\fig{Hubtchi3_tot}), the magnitude of the
transverse susceptibility
in this region tends to increase with increasing $\omega$. This can again be
explained in terms of the DOS of the non-interacting model, which now increases
as the energy is raised (lowered) away from the Fermi energy
up (down) to a van Hove singularity at $3\,$eV ($-3\,$eV) in the
electron-like (hole-like) band. The
``density of excitations'' contributing to the susceptibility therefore
also increases with $\omega$.
As the SDW state is insulating, with a minimum energy of $2\Delta$ required to
excite a quasiparticle
across the gap, the continuum is sharply bounded at
$\omega=2\Delta=42.6\,$meV. The continuum is also bounded
by a dispersing V-shaped feature with minimum at ${\bf q}=0.54\,{\bf
  Q}$, which is not seen for the Hubbard model. The absence of any weight at
small ${\bf q}$ is anticipated from the
band structure in~\fig{EI1_bands}, which shows that the minimum wave vector
for an interband transition with energy $\omega<400\,$meV is ${\bf
  q}\approx0.5\,{\bf Q}$. The V-shaped feature is plotted in detail
in~\fig{interbchi13}(a). As shown in~\fig{interbchi13}(b), it is due to
the weak nesting of the hole band at
${\bf k}=0.23\,{\bf Q}$ with the electron band at ${\bf k}=0.77\,{\bf
  Q}$.
For the energies considered here, to excellent approximation the interband
susceptibility depends only upon
$|\delta{\bf q}|=|{\bf Q}-{\bf q}|$.

%which clearly dominates the transverse susceptibility

%and the resulting interband excitations between these two points

%The weakness of
%this nesting is reflected in the much smaller magnitude of the resonance as
%compared to the paramagnon near ${\bf Q}$.

\begin{figure}
  \includegraphics[width=0.9\columnwidth]{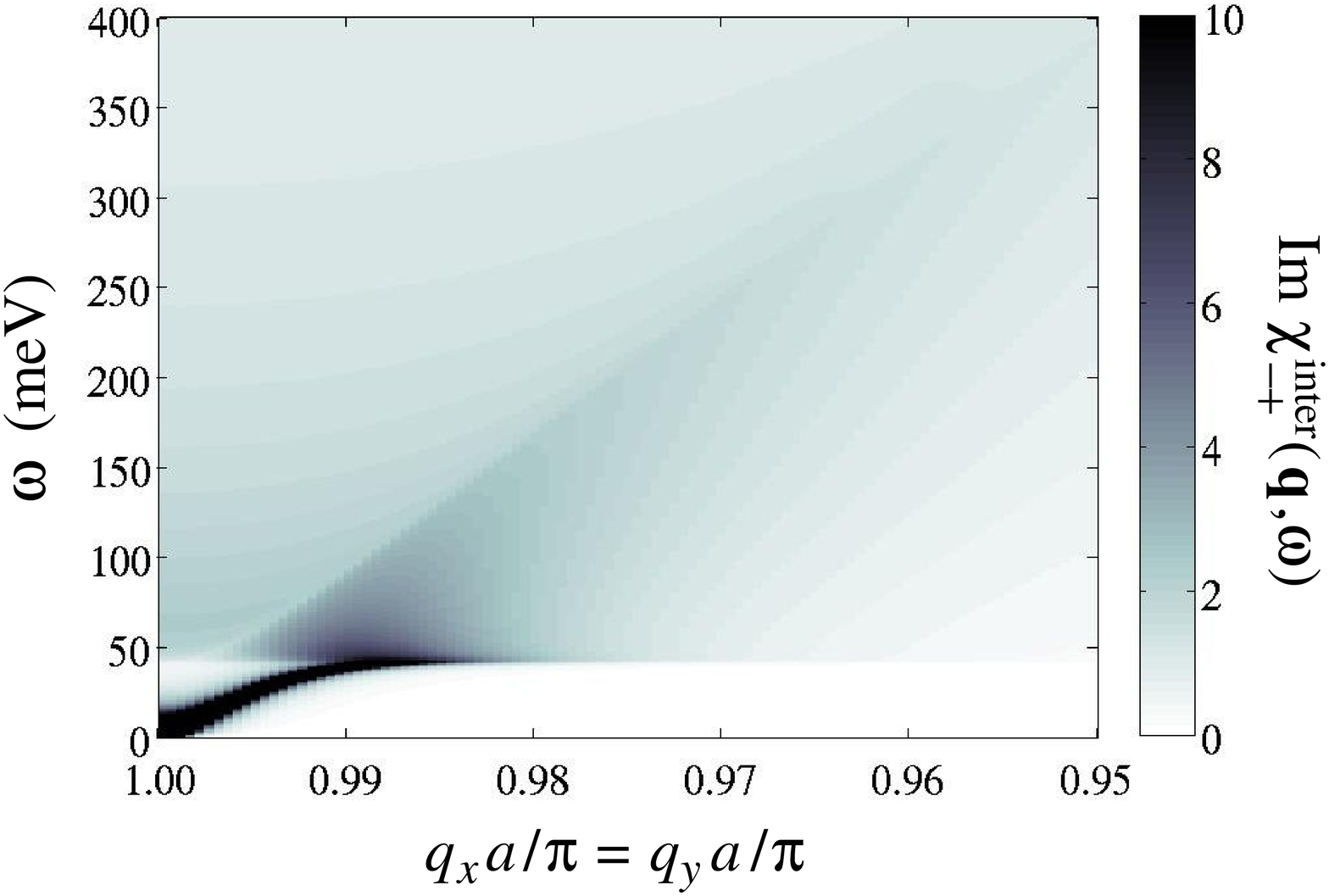}
  \caption{\label{interbchi12lin}(Color online) Imaginary part of the
    interband transverse susceptibility in the insulating excitonic
    model for ${\bf q}=(q_x,q_y=q_x)$ close to ${\bf Q}$. The spin-wave
    dispersion is visible as the thick black line in the lower left-hand
    corner. Note the linear color scale.}
\end{figure}

For ${\bf q}$ near ${\bf Q}$, \fig{perfnest_chi}(a) shows a spin-wave
dispersion which appears to intersect the continuum and continue as a
paramagnon. Figure \ref{interbchi12lin} reveals, however, that the situation is
more
complicated: the spin-wave dispersion does not intersect the continuum, but
instead flattens out as it approaches $\omega=2\Delta$ and disappears at ${\bf
  q}\approx0.985\,{\bf Q}$. As in the Hubbard model, the paramagnon and
spin-wave dispersions appear to avoid one another. The paramagnon
nevertheless seems to connect to significant weight lying just inside the 
continuum region at the intersection point with the spin-wave dispersion.

%which suggests a more complicated relationship between these two features.

We now turn our attention to the intraband contribution to the transverse
susceptibility in~\fig{perfnest_chi}(b). Apart from a forbidden region close
to ${\bf q}=0$, this appears almost
like a mirror image of the interband susceptibility,
albeit much reduced in weight. In particular, we note a
V-shaped dispersing feature at ${\bf q}=0.46\,{\bf Q}$, the tendency of
$\mbox{Im}\,\chi^{\text{intra}}_{-+}({\bf q},\omega)$ to increase with
increasing
$\omega$, and a dispersing feature at the edge of the ${\bf q}\approx0$
forbidden region, which resembles the paramagnon close to ${\bf q}={\bf
  Q}$. The presence of the
V-shaped feature is particularly interesting, as the discussion above
indicates that it is due to interband excitations. Thus we find that
interband excitations give a significant contribution to
the intraband susceptibility. This is confirmed by
examining the Dyson equation for $\chi^{cccc}_{-+,\,00}$, cf.\
\fig{Dyson}(b): for
$g_{cc}=g_{f\!f}=g_\text{2b}=g_\text{2a}=0$, as assumed here, the intraband
susceptibilities do
not appear on the right-hand side of the equation so that the RPA-enhancement of
$\chi^{cccc}_{-+,\,00}$ stems only from the umklapp susceptibilities
$\chi^{cfcc}_{-+,\,{\bf Q}0}$ and 
$\chi^{fccc}_{-+,\,{\bf Q}0}$. The coupling to these terms in the Dyson equations
is through the
anomalous Green's functions $G^{cf}$ and
$G^{fc}$, which reflect
the mixing of the states in the electron-like and hole-like bands separated by
the nesting vector ${\bf Q}$ in the SDW phase.
Consequently, the intraband susceptibility is
similar to the interband susceptibility, but shifted by ${\bf Q}$.

%These anomalous Green's functions hence allow an intraband excitation with
%wave vector ${\bf q}$ to become an interband excitation with wave vector ${\bf
%q}+{\bf Q}$, and \emph{vice versa}.

The total transverse susceptibility
in~\fig{perfnest_chi}(c) clearly shows the partial symmetry of the
response about ${\bf q}={\bf Q}/2$ but also the asymmetric
distribution of weight. $\mbox{Im}\,\chi_{-+}({\bf q},\omega)$ for $|{\bf
q}|<|{\bf Q}|/2$ is roughly one order of magnitude smaller than at ${\bf
  q}' = {\bf Q}-{\bf q}$.

%The much weaker
%feature in the intraband susceptibility [\fig{perfnest_chi}(b)] at ${\bf
%  q}=0.46{\bf Q}$ has an identical explanation, although here the resonance is
%shifted to smaller ${\bf q}$ by the Umklapp processes.

\begin{figure}
  \includegraphics[width=0.9\columnwidth,clip]{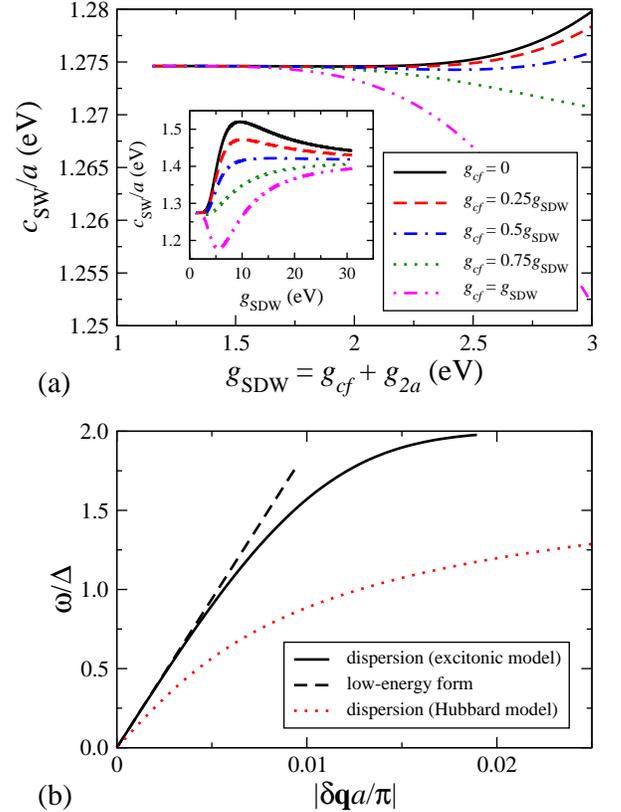}
  \caption{\label{EI1sw}(Color online) (a) Spin-wave velocity $c_{\text{SW}}$
    as a
    function of $g_{\text{SDW}}=g_{cf}+g_\text{2a}$ in the insulating excitonic
    model. Inset: $c_{\text{SW}}$ for a larger range of
    $g_{\text{SDW}}$.
    (b) Spin-wave dispersion in the excitonic model and its low-energy linear
    form for $g_{\text{SDW}}=1.8\,$eV as a function of $\delta{\bf
    q}={\bf q}-{\bf
    Q}$. Shown for comparison is the spin-wave dispersion in the Hubbard
    model from~\fig{Hubsw}(b) for the same $T=0$ gap.}
\end{figure}

\subsubsection{Spin-wave velocity}

The calculation of the spin-wave velocity proceeds as for the Hubbard
model. For $g_{cc}=g_{f\!f}=g_\text{2b}=0$, solving the Dyson equations for
the interband susceptibilties in~\eq{eq:EI:chixxcf} and~\eq{eq:EI:chixxfc}
again involves the inversion of a $2\times2$ matrix, which has the determinant
\beqarray
{\cal D}({\bf q},\omega) & = &
\left[1-g_{cf}\chi^{cf\!fc\,(0)}_{-+,\,00}-g_\text{2a}\chi^{cfcf\,(0)}_{-+,00}
\right]
\notag \\
&& \quad{}\times
\left[1-g_{cf}\chi^{fccf\,(0)}_{-+,\,00}-g_\text{2a}\chi^{fcfc\,(0)}_{-+,00}
\right]
\notag \\
&& {}-
\left[g_{cf}\chi^{fcfc\,(0)}_{-+,\,00}+g_\text{2a}\chi^{fccf\,(0)}_{-+,\,00}
\right]
\notag \\
&&
\quad{}\times\left[g_{cf}\chi^{cfcf\,(0)}_{-+,\,00}+g_\text{2a}\chi^{cf\!fc\,(0)
} _
{-+,\, 00} \right] .
\label{eq:EI:Dqomega}
\eeqarray
Expanding this determinant about $\omega=0$ and $\delta{\bf q} ={\bf
Q}-{\bf q}=0$, we obtain the low-energy linear form
$\omega=c_{\text{SW}}\,|\delta{\bf q}|$
of the spin-wave dispersion. For the band structure considered here,
the spin-wave velocity is given by
\beq
c_{\text{SW}} = \sqrt{\frac{2t^2a_3[a_0(g^2_\text{2a}-g_{cf}^2) -
g_\text{2a}]}{(a_1^2 +
    2a_0a_2)(g_\text{2a}^2 - g_{cf}^2) - 2a_2g_\text{2a}}}  , \label{eq:EI:csw}
\eeq
where
\beq
a_{0} = \frac{1}{4V}\sum_{\bf k}\frac{\Delta^2}{E_{{\bf k}}^3}, \qquad
a_{1} = \frac{1}{4V}\sum_{\bf k}\frac{\epsilon^{c}_{{\bf k}+{\bf Q}}}{E_{{\bf k}}^3} ,
\eeq\beq
a_{2} = \frac{1}{8V}\sum_{\bf k}\frac{1}{E_{{\bf k}}^3} ,
\eeq
\beqarray
a_3 &=& \frac{1}{2V}\sum_{\bf k}\left\{\frac{\epsilon^{c}_{{\bf k}+{\bf Q}}}{4E^{3}_{\bf k}}t\cos{k_xa} \right. \notag \\
&&\left. + \frac{2\Delta^4 +
  \Delta^2(\epsilon^{c}_{{\bf k}+{\bf Q}})^2 - (\epsilon^{c}_{{\bf k}+{\bf Q}})^4}{E_{{\bf k}}^5}\,\sin^{2}k_xa\right\} ,
\eeqarray
and
\beq
E_{\bf k}=\sqrt{(\epsilon^{c}_{{\bf k}+{\bf Q}})^2 + \Delta^2} .
\eeq
We plot $c_{\text{SW}}$ as a function of $g_{\text{SDW}}$ for different values
of $g_{cf}$ in~\fig{EI1sw}(a). The behaviour of the spin-wave velocity for
$g_\text{SDW}\gg{t}$ is included as an inset. For $g_{\text{SDW}}\geq g_{cf}$
we always find $c^2_{\text{SW}}>0$; for sufficiently large 
$g_{cf}>g_{\text{SDW}}$, however, we have $c^2_{\text{SW}}<0$ which indicates that
the system becomes unstable towards a different ground state. This is not
surprising, as 
for $g_{cf}>g_{\text{SDW}}>0>g_{2a}$ the effective coupling $g_{\text{SDW}}$
is smaller than that for the CDW or SCDW. In the opposite case
$g_{2a}>g_{\text{SDW}}>0>g_{cf}$ the coupling constants for the SDW and CCDW
states are equal and always greater than those for the CDW and SCDW, and so
the SDW remains stable.
In~\fig{EI1sw}(b) we plot the
spin-wave dispersion in the excitonic model as a function of $\delta{\bf
  q}$. Compared to a Hubbard model with identical $T=0$ gap, the spin-wave
dispersion has both a higher 
low-energy velocity and remains approximately linear up to higher
energies [see~\fig{Hubsw}(b)].

Although~\eq{eq:EI:csw} is a rather complicated function of $g_{cf}$ and
$g_{2a}$, for $g_{\text{SDW}}<3\,$eV the spin-wave velocity in the excitonic
model shows remarkably little dependence upon the interaction constants, in
contrast to the Hubbard model results in~\fig{Hubsw}(a). Instead, the value of
$c_\text{SW}$ is fixed by the band structure: for an excitonic gap
$\Delta\ll{t}$ (the weak-coupling limit) we have to excellent approximation
$c_\text{SW}\approx\widetilde{v}_{F}/\sqrt{2}$ where $\widetilde{v}_F$ is the
average Fermi velocity. This is anticipated by the results of
Refs.~\onlinecite{Fedders1966} and \onlinecite{Liu1970} for chromium, where 
it was found that $c_{\text{SW}}=\sqrt{{v}_{e}{v}_{h}/3}$
where $v_{e(h)}$ is the electron
(hole) Fermi velocity, and the factor of $1/\sqrt{3}$ arises because a
three-dimensional Fermi surface is considered. It is
also consistent with our observation that
$\chi_{-+}({\bf q},\omega)$ is rather insensitive to the choice of $g_{cf}$
and $g_\text{2a}$ for small $g_{\text{SDW}}$. 

The behaviour of $c_\text{SW}$ in the strong-coupling regime of the Hubbard
and excitonic models is also qualitatively different. In the
former, the $U\gg{t}$ limit of~\eq{eq:Hub:csw} gives $c_{\text{SW}}=\sqrt{2}J$
where $J=4t^2/U$ is the exchange integral of the corresponding effective
Heisenberg model.~\cite{Schrieffer1989,Singh1990,Chubukov1992} In the
excitonic model, however,
the inset of~\fig{EI1sw}(a) reveals that $c_{\text{SW}}$ has only weak
dependence upon the interaction strength for $g_{\text{SDW}}\gg{t}$. A
strong-coupling expansion of~\eq{eq:EI:csw} gives the limiting value
$c_{\text{SW}}=\sqrt{2}ta$. The interpretation of the strong-coupling limit in
the excitonic model is not straightforward: as
$g_\text{SDW}\rightarrow\infty$, simultaneous occupation of the $c$ and $f$
states on the same site is forbidden, but double occupation of the $c$ and $f$
states is allowed. Since we work at half-filling, one might expect a
checkerboard orbital ordering with filled $c$ states on one sublattice and
filled $f$ states on the other, which is incompatible with SDW
order. However, it has been shown in a \emph{spinless} two-band model that
such a state is unstable towards an excitonic insulator or a phase with either
the $c$ or $f$ states fully occupied for 
$\epsilon_{0}\neq0$.~\cite{Batista2002} How this result would change in the
presence of spin is not clear. In any case, the 
$g_{\text{SDW}}\gg{t}$ limit seems somewhat unphysical without also considering
$g_{cc}$ and $g_{ff}$ to be large, and so we do not further discuss the
strong-coupling regime here.

%The dispersion significantly flattens as it approaches
%the edge of the continuum excitation region at  
%$\omega=2\Delta$, and disappears entirely before reaching $|\delta{\bf
%  q}a/\pi|=0.02$. As mentioned above, it is not clear if the spin-wave
%dispersion persists into the continuum region as the paramagnon feature shown
%in~\fig{interbchi12lin}.

The evaluation of the intraband susceptibilities proceeds similarly
but here the denominator is ${\cal D}({\bf q}+{\bf Q},\omega)$. This
yields an identical spin-wave dispersion but shifted to ${\bf q}=0$. As in
the Hubbard model, however, the spin wave has vanishing weight close to the
zone center and is barely visible as it exits the continuum in the lower
right-hand corner of~\fig{perfnest_chi}(b).

\subsection{Metallic SDW state}\label{subsec:imperf}

\begin{figure}
  \includegraphics[width=\columnwidth,clip]{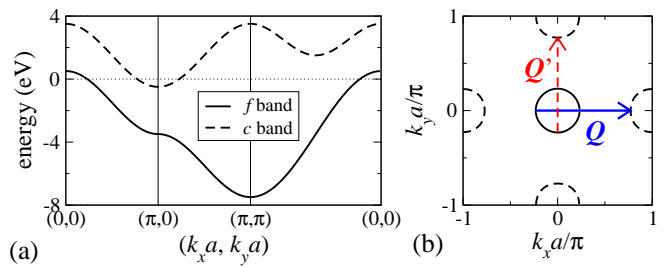}
  \caption{\label{EI2_bands}(Color online) (a) Band structure and (b)
    Fermi surface of the non-interacting metallic excitonic model. In
    (b), the nesting vectors ${\bf Q}=(\pi/a,0)$ and
    ${\bf Q}'=(0,\pi/a)$ are also shown.}
\end{figure}

It is more generally the case that the nesting condition $\epsilon^{c}_{\bf
    k}\approx-\epsilon^{f}_{{\bf k}+{\bf Q}}$ is only approximately satisfied.
Furthermore, there may be portions of
the Fermi surface that do not participate in the excitonic instability, as
is the case in chromium,\cite{FawcettCr,Kulikov1984,Rice1970} The
pnictides
also have a complicated Fermi surface involving several
bands. Although the numerous models for the band
structure differ in their
details,\cite{Singh2008,Korshunov2008,Kuroki2008,Raghu2008,%
Lorenzana2008,Ran2009,Yu2009,Brydon2009}
there is general agreement that in the ``unfolded'' Brillouin zone corresponding
to the 2D iron sublattice the nesting of
hole pockets at $\mathbf{k}=(0,0)$ with electron pockets at $(\pi/a,0)$ or
$(0,\pi/a)$ is primarily responsible for the SDW.
In the physical, tetragonal Brillouin zone, both $(\pi/a,0)$ and $(0,\pi/a)$
are folded back onto the M point, leading to two
electron pockets around that point.\cite{Yu2009,Brydon2009} The wave vectors
in the present paper refer to the unfolded zone.
Apparently only one of the
electron pockets undergoes the excitonic instability, yielding a SDW with
ordering vector ${\bf Q}=(\pi/a,0)$, say.
The other electron pocket at ${\bf 
  Q}'=(0,\pi/a)$ remains ungapped. We can capture the basic features of this
scenario within a two-band model by including one hole pocket around $(0,0)$
and treating the two electron pockets as belonging to the same band. We thus
assume the band structure
\begin{subequations}
\label{eq:EI:imperfbs}
\beqarray
\epsilon^{c}_{\bf k} & = & 2t\cos{k_xa}\cos{k_ya} +
\epsilon_{c}, \\
\epsilon^{f}_{\bf k} & = & 2t\left(\cos{k_xa} + \cos{k_ya}\right) +
\epsilon_{f}.
\eeqarray
\end{subequations}
We take $t=1\,$eV, $\epsilon_c=1.5\,$eV, $\epsilon_{f}=-3.5\,$eV, and fix the
doping
at $n=1.916$, which gives electron and hole pockets of identical area. The
band structure and Fermi surface are illustrated in~\fig{EI2_bands}(a) and (b),
respectively. Note that the hole pocket is nearly but not quite perfectly
nested with both electron pockets. We impose a single-$\mathbf{Q}$ SDW with
ordering
vector ${\bf Q}=(\pi/a,0)$. For $g_{\text{SDW}}=1.873$eV we find a mean-field
state with
critial temperature $T_{\text{SDW}}=132\,$K and $T=0$ gap $\Delta=21.3\,$meV. In
the $T=0$ SDW state both the hole pocket at the zone center and the electron
pocket at $(\pi/a,0)$ are completely gapped, while the electron pocket at
$(0,\pi/a)$ remains intact.

%(solid blue arrow)(dashed red arrow)

\begin{figure*}
  \includegraphics[width=1.4\columnwidth]{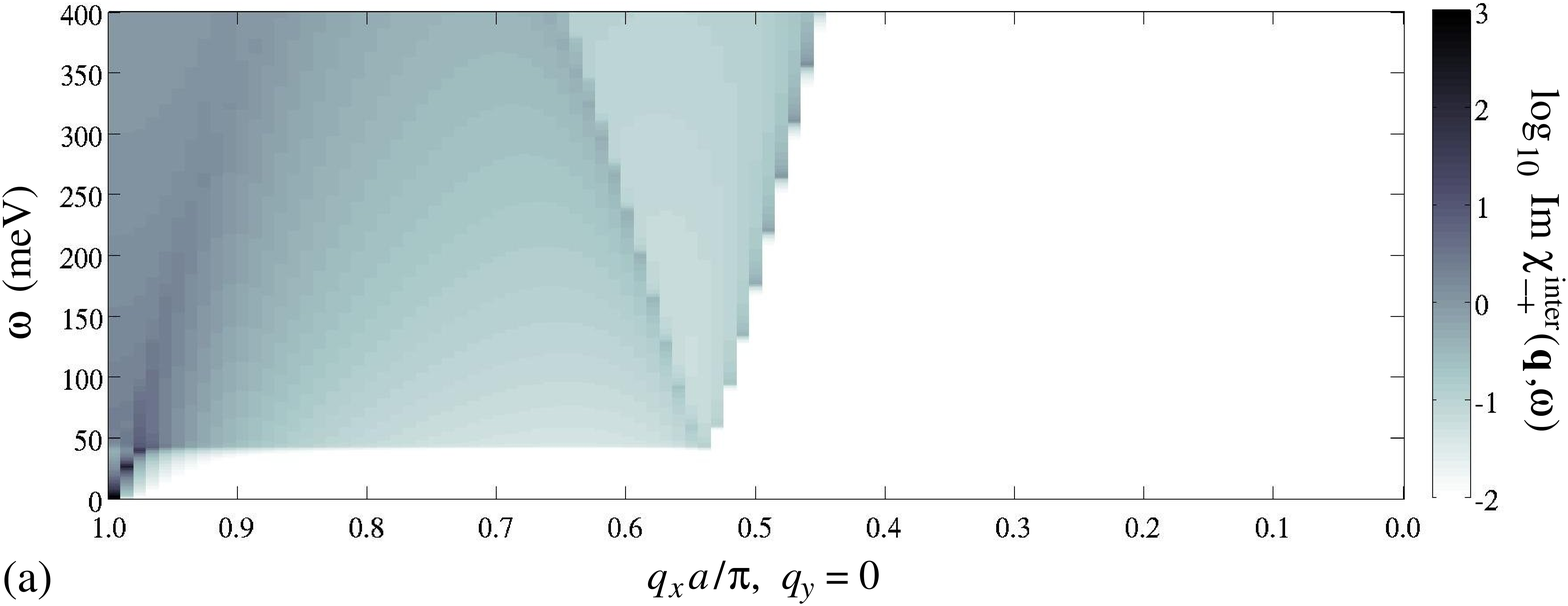}
  \includegraphics[width=1.4\columnwidth]{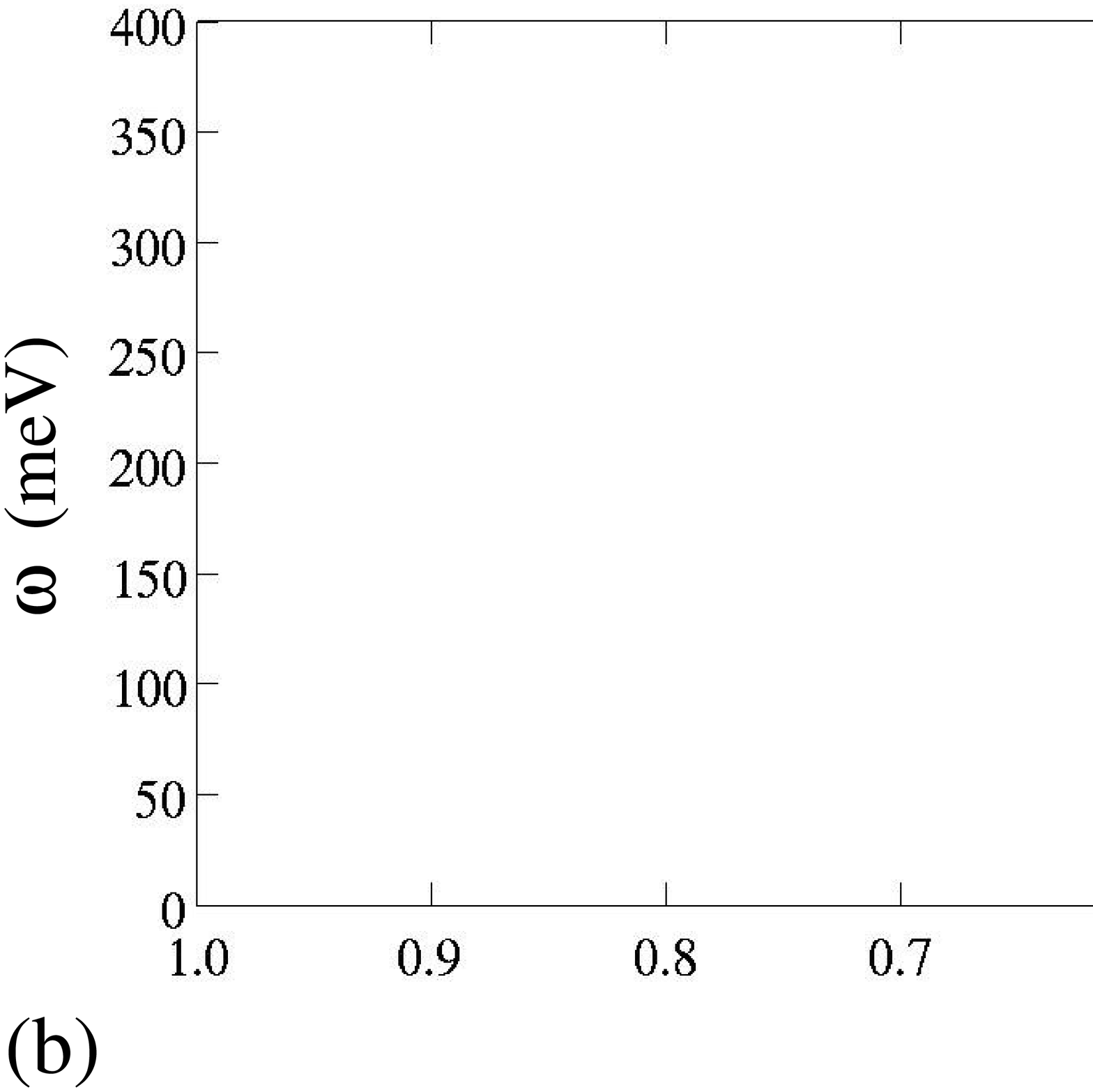}
  \includegraphics[width=1.4\columnwidth]{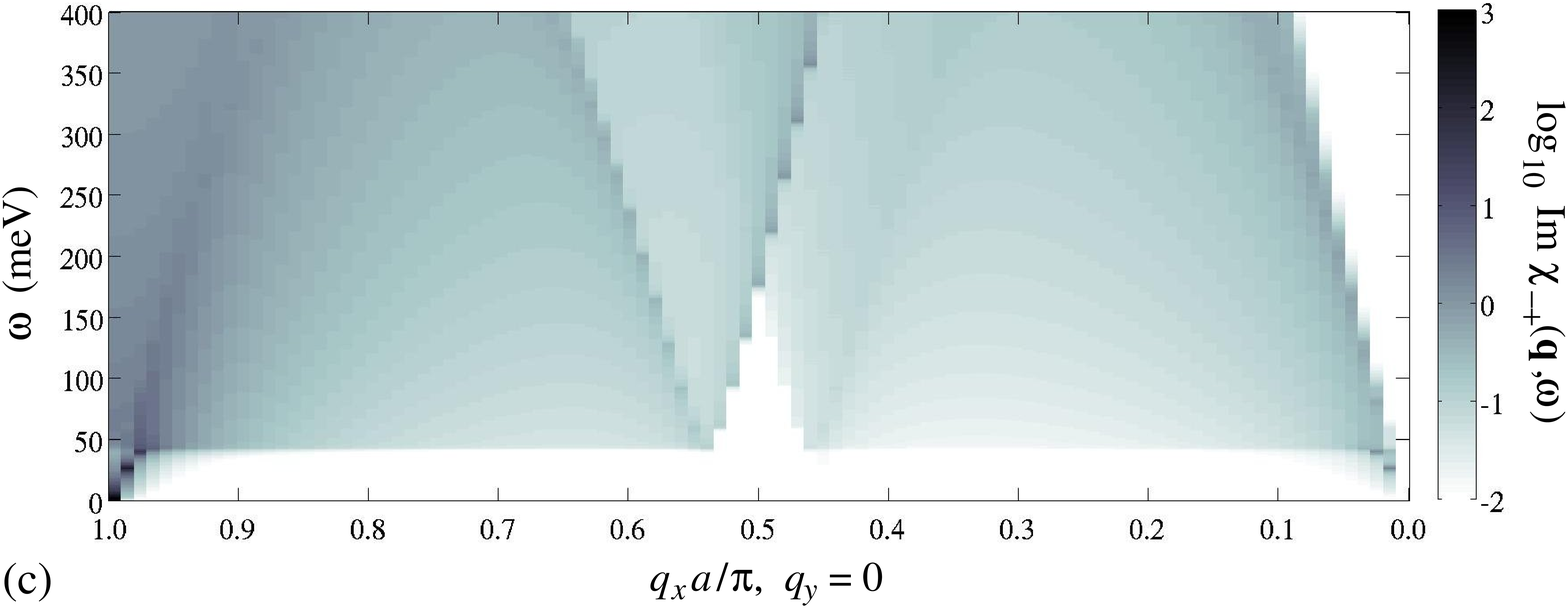}
  \caption{\label{imperfnest_chi1}(Color online) Imaginary part of the (a)
    interband, (b) intraband, and (c) total transverse
  susceptibility in the metallic excitonic model for ${\bf
    q}=(q_x,0)$. The spin-wave dispersion is visible as the dark
  line at $\omega<2\Delta$ close to ${\bf Q}$ in (a) and (c). Note
  the logarithmic color scale.}
\end{figure*}

\begin{figure*}
  \includegraphics[width=1.4\columnwidth]{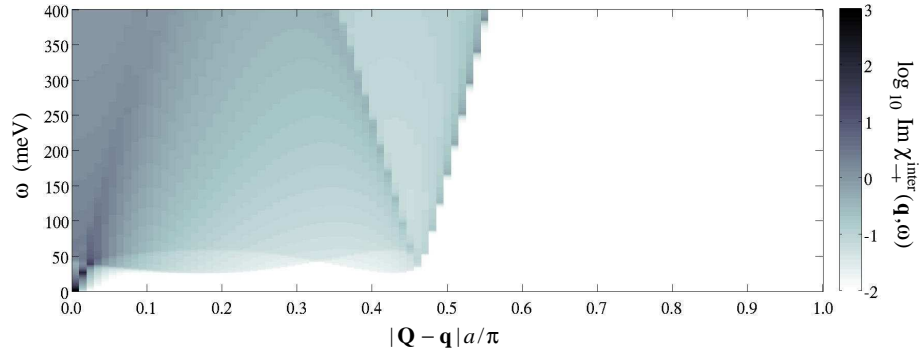}
  \caption{\label{imperfnest_chi1b}(Color online) Imaginary part of the
    interband transverse
  susceptibility in the metallic excitonic model for ${\bf
    q}=(\pi/a-\widetilde{q}\cos\theta,\widetilde{q}\sin\theta)$ for
  $\theta=\pi/8$. The spin-wave dispersion is visible as the dark
  line at $\omega<2\Delta$ close to ${\bf Q}$. Note
  the logarithmic color scale.}
\end{figure*}

% XXX Fig. 13 needs to be recreated from jpeg or source! XXX

The imaginary parts of the interband, intraband, and total transverse
susceptibilities for ${\bf q}=(q_x,0)$ are shown in~\fig{imperfnest_chi1}(a),
(b), and (c), respectively. Our results are very similar to those for
the insulating SDW model in~\fig{perfnest_chi}.
The slightly higher magnitude of the transverse susceptibility is due to
the greater density of states in the 
electron-like band.
The similarity is not surprising, as the relevant
excitations in both models have identical origin, i.e., excitations between
states close to two Fermi pockets which are gapped by an excitonic SDW
instability. The states close to the ungapped Fermi pocket do not
contribute to the interband susceptibility for the plotted range of $({\bf
  q},\omega)$. Although these states do contribute to the intraband
susceptibility for 
small values of ${\bf q}$, they are only negligibly mixed with states in the
hole-like band,
and thus are not RPA-enhanced by the interband interactions.

%; the location of the V-shaped feature at a similar fraction of ${\bf Q}$ as in
%the perfectly-nested model is a consequence of the band structure. 

%We find that the contribution to the transverse susceptibility of
%such intraband excitations is dwarfed by the contribution of those near the
%gapped Fermi pockets. 

\begin{figure}
  \includegraphics[width=0.9\columnwidth,clip]{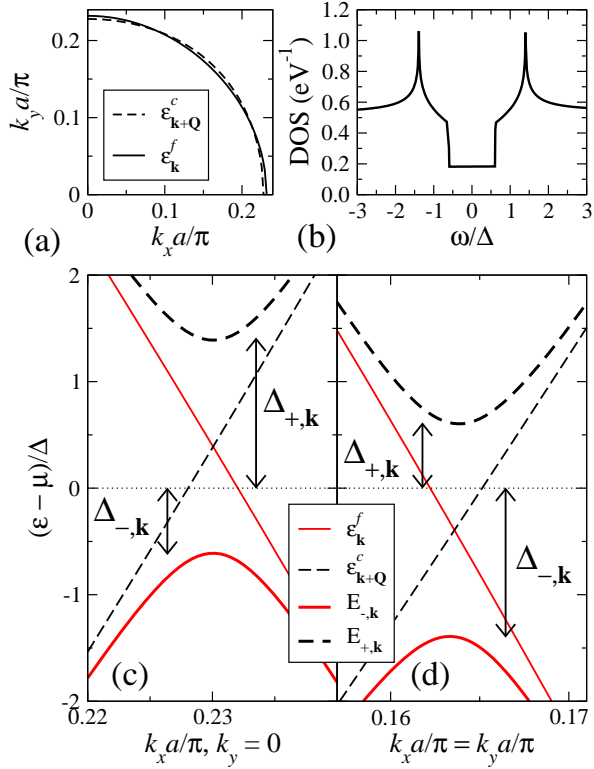}
  \caption{\label{deltapm}(Color online) (a) Electron and hole Fermi pockets
    superimposed upon one another. (b) Density of states close to the Fermi
    energy. (c) Comparison of the band structure near the intersection of
    $\epsilon^{c}_{{\bf k}+{\bf Q}}$ and $\epsilon^{f}_{{\bf k}}$ in the
    normal and SDW phases for ${\bf k}=(k_x,0)$. (d) Same as in (c), but for
    ${\bf k}=(k_x,k_y=k_x)$.}
\end{figure}

In contrast to the insulating SDW state studied in~\Sec{subsec:perf}, here the
interband
susceptibility does not just depend upon $|\delta{\bf q}|$:
although it is identical for ${\bf q}=(q_x,0)$ and
$(\pi/a,\pi/a-q_x)$ by tetragonal symmetry, and quantitatively very similar
along ${\bf q}=(\pi/a-q_x/\sqrt{2},{\pm}q_x/\sqrt{2})$, away from these
high-symmetry lines in ${\bf q}$-space we find that the continuum can extend
to significantly lower 
energies. This is shown in~\fig{imperfnest_chi1b}, where we plot
$\text{Im}\,\chi^{\text{inter}}_{-+}({\bf q},\omega)$ for ${\bf
  q}=(\pi/a-\widetilde{q}\cos\theta,\widetilde{q}\sin\theta)$ with
$\theta=\pi/8$. Although the response for $\omega>100\,$meV is very similar to
that in~\fig{imperfnest_chi1}(a), we see that the lower edge of the continuum
is not constant at $\omega=2\Delta$, but instead shows higher and lower
thresholds which coincide only at special values of $\mathbf{q}$.

%$|{\bf Q} - {\bf q}|=0$, $0.33\,\pi/a$, and $0.46\,\pi/a$. 

The origin of this threshold behavior is the imperfect
nesting of the Fermi pockets. Consider~\fig{deltapm}(a), which shows the
superimposed hole and back-folded electron Fermi pockets:
along ${\bf k}=(k_x,0)$ or $(0,k_y)$, the width of the hole Fermi
pocket is greater than that of the electron one, while the reverse is true for
${\bf k}=(k_x,k_y=k_x)$ or $(k_x,k_y=-k_x)$. In the former case, the
intersection of the non-interacting electron-like and hole-like dispersions
therefore occurs above the Fermi energy [\fig{deltapm}(c)], while in the
latter it occurs below the Fermi energy [\fig{deltapm}(d)]. In
the reconstructed bands of the excitonic model, the SDW
gap is always centered at the point of intersection of the original bands,
as can be seen from~\eq{eq:EI:rebuilt} and in Figs.\ \ref{deltapm}(c) and (d).
%, which
%correspond to the  
%limits of the anisotropy of $\Delta_{+,{\bf k}}$ and $\Delta_{-,{\bf
%    k}}$. 
In general, the difference between the Fermi energy and the bottom of the
reconstructed
electron-like band, $\Delta_{+,{\bf k}}$, and the difference between the Fermi
energy and the top of the reconstructed hole-like band,
$\Delta_{-,{\bf k}}$, will be unequal.
%The distribution in the extrema of the
%band edges is seen in~\fig{deltapm}(b). 
The minimum energy for an
interband excitation with wave vector ${\bf Q}+\delta{\bf q}$ is therefore
$\min_{{\bf k}}(\Delta_{\pm,{\bf k}}+\Delta_{\mp,{\bf k}+\delta{\bf q}})$. For
$\delta{\bf q}$ along the high-symmetry 
directions mentioned above, the tetragonal symmetry of the Fermi pockets
ensures that this minimum energy is $2\Delta$. Away from these
directions, however, the energy difference depends upon
$\delta{\bf q}$. For example, the states near the Fermi surface
in~\fig{deltapm}(a)
at ${\bf k}=(0,k_y)$ and $(k_x,k_y=k_x)$ are connected by $\delta{\bf q} =
(\widetilde{q}\cos({\pi}/{8}),-\widetilde{q}\sin({\pi}/{8}))$
with $\widetilde{q}=0.17\,\pi/a$; from~\fig{deltapm}(c) and (d) we see that the
minimum energy for single-particle excitations with this wave vector is
$1.2\,\Delta=25.6\,$meV, which marks the lowest edge of the continuum
in~\fig{imperfnest_chi1b}. The upper threshold originates from the maximum
energy connecting the top of the hole-like band and the
bottom of the electron-like band, which for this ${\bf q}$ is
$2.8\,\Delta=59.6\,$meV. For the
remainder of this paper we shall restrict ourselves to
high-symmetry directions.

\begin{figure*}
  \includegraphics[width=1.4\columnwidth]{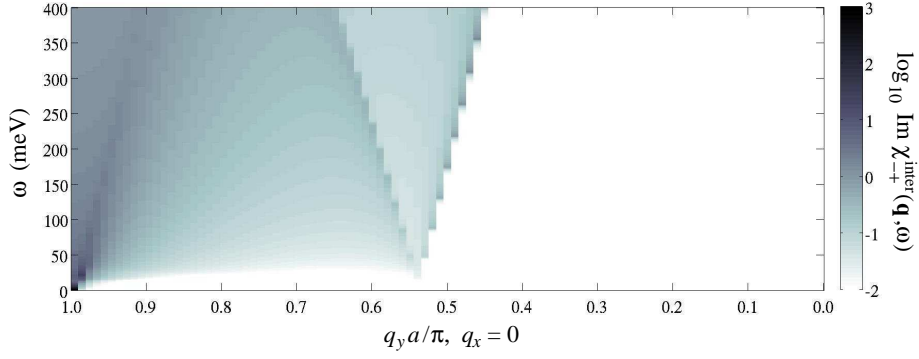}
  \caption{\label{imperfnest_chi2}(Color online) Imaginary part of the
    interband transverse
  susceptibility in the metallic excitonic model for ${\bf
    q}=(0,q_y)$. The spin-wave dispersion is visible as the dark
  line at $\omega<2\Delta$ close to ${\bf Q}'$ ($q_y=0$). Note
  the logarithmic color scale.}
\end{figure*}

% discuss contribution of ungapped portion of Fermi surface to interband
% susceptibility near Q_2

Since only one electron pocket is gapped, the directions $(q_x,0)$ and
$(0,q_y)$ are not equivalent.
The imaginary part of the interband transverse susceptibility along ${\bf
  q}=(0,q_y)$ is shown in~\fig{imperfnest_chi2}. This is the direction towards
the second nesting vector $\mathbf{Q}'=(0,\pi/a)$, which was not selected by the
SDW instability. For ${\bf q}$ sufficiently
close to ${\bf Q}'$, we thus find the response generated by transitions
between states near the (gapped) hole Fermi pocket and states near the
ungapped electron Fermi pocket. At $\omega\gtrsim 2\Delta$,
this is very similar to the interband susceptibility near
${\bf Q}$ [\fig{imperfnest_chi1}(a)], reflecting the small changes to the band
structure at high energies upon opening of the SDW gap.
The differences are more striking at lower energies.
In particular, comparing \fig{imperfnest_chi2} to \fig{imperfnest_chi1}(a),
we see that the continuum extends to lower energies
close to ${\bf Q}'$ than close to ${\bf Q}$. The minimum energy
required for a single-particle excitation
between the states near the ungapped electron pocket and the gapped
hole pocket is smaller than $2\Delta$, thus giving a lower threshold
for the continuum near ${\bf Q}'$.

%between the interband susceptibility close to ${\bf Q}$ and ${\bf Q}'$

%where the presence and absence of the SDW gap, respectively, becomes important

\begin{figure}
  \includegraphics[width=0.9\columnwidth]{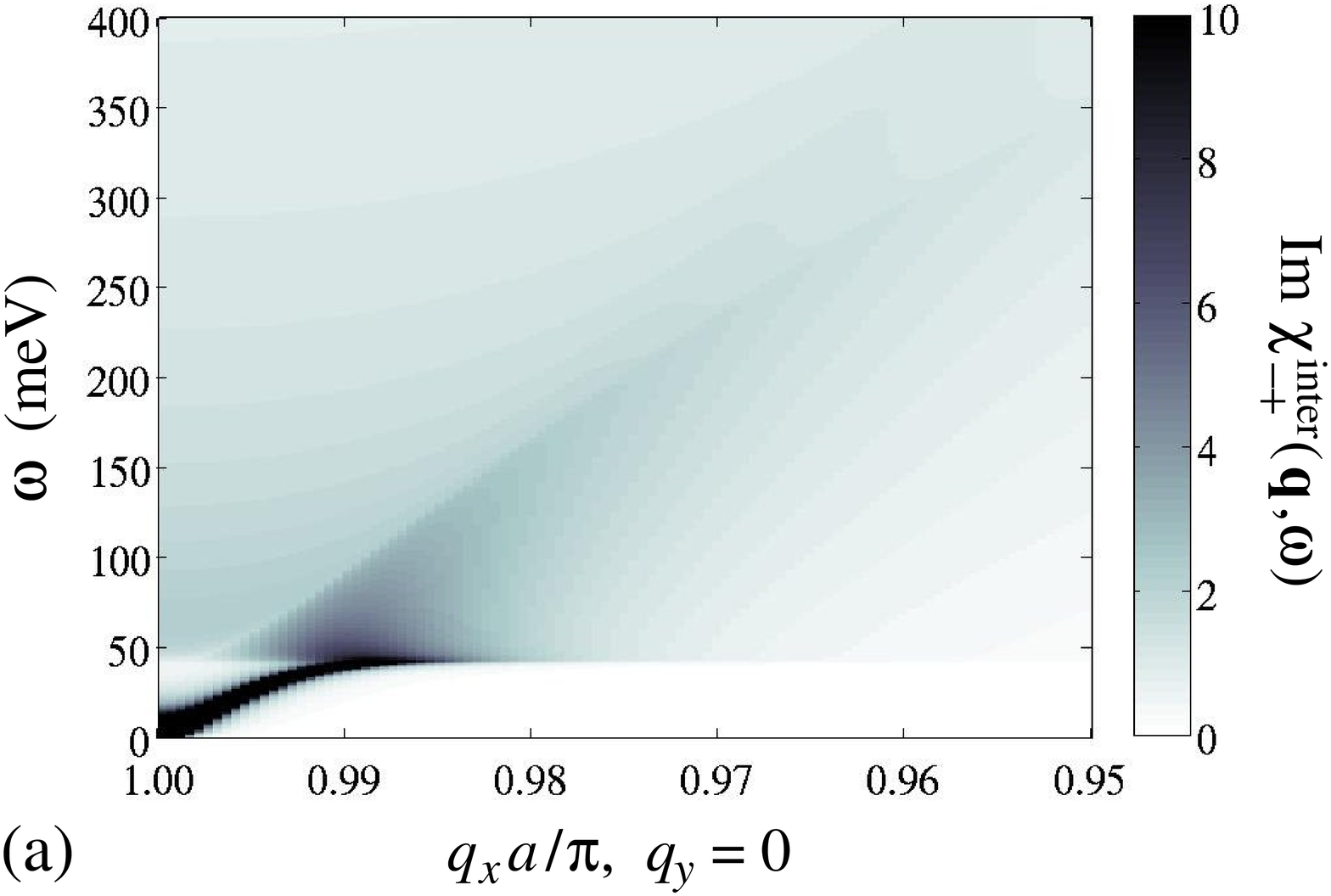}
  \includegraphics[width=0.9\columnwidth]{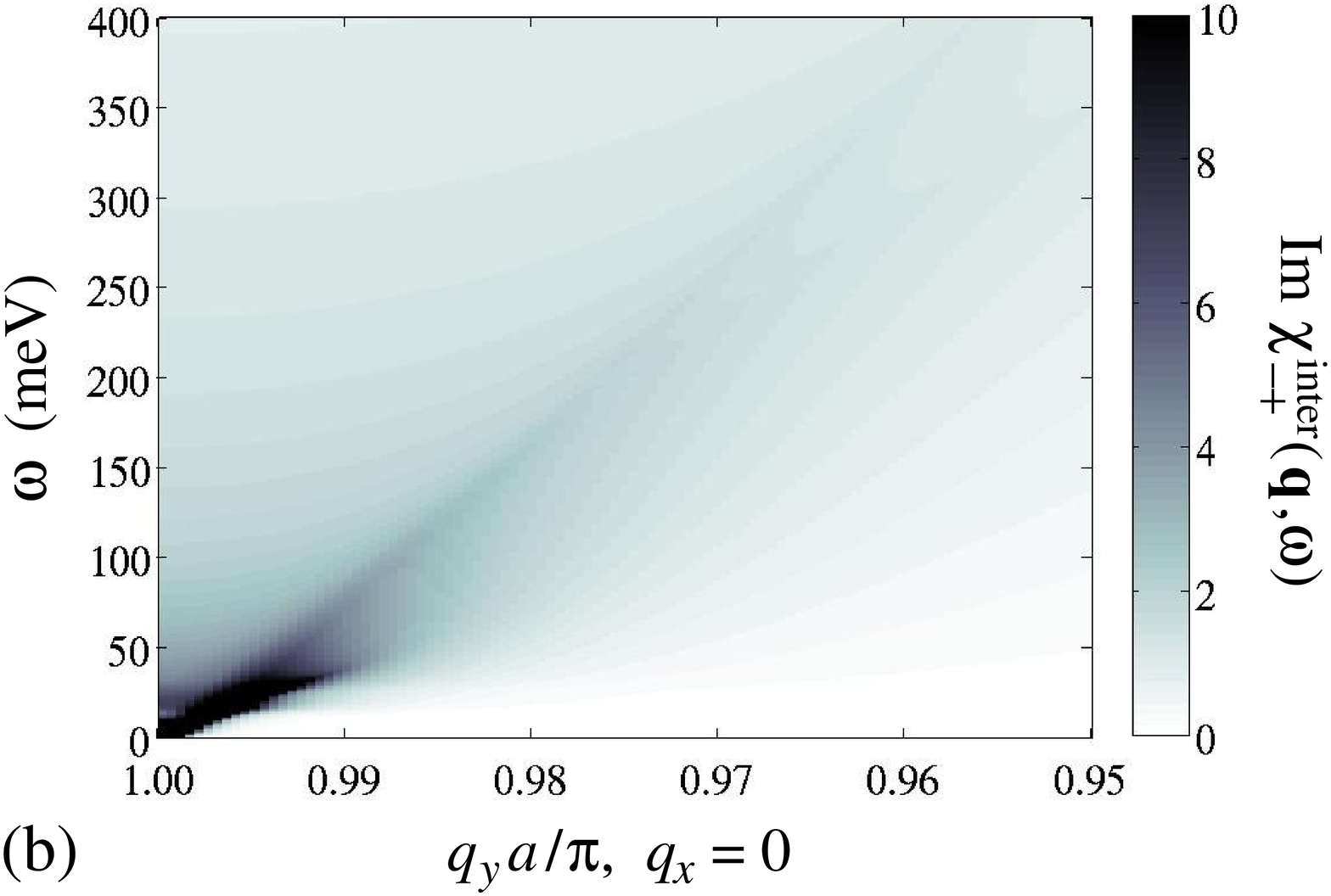}
  \caption{\label{interbchi6_1vs3}(Color online) Imaginary part of the
    interband transverse susceptibility in the metallic excitonic
    model for (a) ${\bf q}=(q_x,0)$ close to ${\bf Q}$ and (b) ${\bf
      q}=(0,q_y)$ close to ${\bf Q}'$. In both panels the spin-wave
    dispersion is visible as the thick black line in the bottom left-hand
    corner. Note that in (b) that the continuum region starts at
    $\omega\approx0.6\,\Delta$. In both panels we use a linear color scale.}
\end{figure}

\begin{figure}
  \includegraphics[width=0.9\columnwidth,clip]{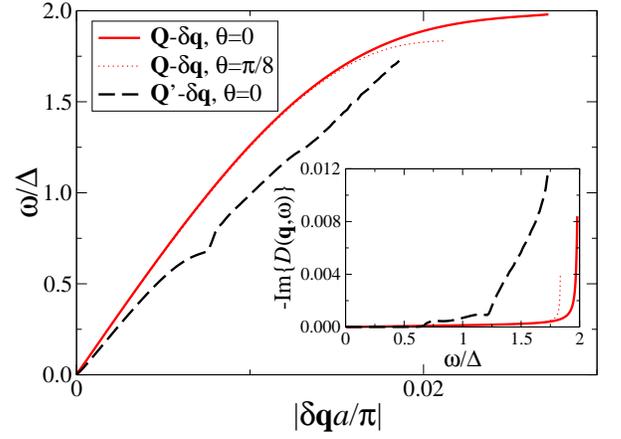}
  \caption{\label{interbchi6_sw}(Color online) Spin-wave
    dispersion in the metallic excitonic model close to ${\bf Q}$ for
    $\delta{\bf q} =
    (-\widetilde{q}\cos\theta,\widetilde{q}\sin\theta)$ with $\theta=0$
    (solid red line) and $\theta=\pi/8$ (thin dotted red line). The
    dispersion for $\theta=\pi/4$ is indistinguishable from the $\theta=0$ case.
    We also plot the spin-wave dispersion close to ${\bf Q}'$ (dashed
    black line). Inset: imaginary part of ${\cal D}({\bf
      q},\omega)$ for each dispersion. We have used a
    20000$\times$20000 ${\bf k}$-point mesh and a width $\delta=0.1\,$meV to
    calculate the mean-field susceptibilities.
    Note that the finite value of
    $\text{Im}\,{\cal D}({\bf q},\omega)$ for the
    spin waves close to ${\bf Q}$ for $\omega<2\Delta$
    is an artifact of the finite width $\delta$.}
\end{figure}

% spin-wave dispersion

Another significant difference concerns the
spin-wave dispersion. The spin-wave
dispersion near ${\bf Q}$ is visible in the lower left-hand corner
of~\fig{imperfnest_chi1}(a), and it intersects the continuum and appears to
continue as a paramagnon. From~\fig{imperfnest_chi2}, we see that
there is also a gapless Goldstone mode at ${\bf q}={\bf Q}'$.
This mode is gapless since it rotates the single-$\mathbf{Q}$ SDW into a
superposition of $\mathbf{Q}$ and $\mathbf{Q}'$ SDWs, which is degenerate with
the single-$\mathbf{Q}$ SDW in our tetragonal model.
Although there
appears to be a spin-wave branch around ${\bf Q}'$, it is not as distinct as
in~\fig{imperfnest_chi1}(a) due to the lower threshold of the continuum. We
therefore plot the interband transverse susceptbilities for a finer ${\bf
  q}$-resolution near ${\bf Q}$ and ${\bf Q}'$ in~\fig{interbchi6_1vs3}(a)
and (b), respectively. As expected from the discussion above, the former is
qualitatively identical to~\fig{interbchi12lin}. The latter, in contrast,
shows several novel features: the spin-wave dispersion
does not curve away from the edge of the continuum but rather
intersects it with little change in velocity and the spin-wave
and paramagnon features approach much closer to one another than for ${\bf
  q}\approx{\bf Q}$. Although it is not clear from~\fig{interbchi6_1vs3}(b),
the spin-wave and paramagnon dispersions do not intersect, and the spin waves
become damped at $\omega\approx1.7\Delta$.

%which originates from the degeneracy of the SDW states with ordering vector
%${\bf Q}$ and ${\bf Q}'$ in our tetragonal model.

To obtain the spin-wave dispersion, we must again solve
$\text{Re}\,{\cal D}({\bf q},\omega)=0$ with ${\cal D}({\bf q},\omega)$
given
by~\eq{eq:EI:Dqomega}. We have not been able to obtain analytical expressions
for the spin-wave velocity, however, as the Fermi distribution functions
appearing in the mean-field susceptibilities cannot be expanded as a Taylor
series in $\delta{\bf q}$ due to the ungapped electron Fermi pocket.
Plotting the dispersions at ${\bf q}\approx{\bf Q}$ and
${\bf q}\approx{\bf Q}'$ in~\fig{interbchi6_sw}, we see that the velocity at
${\bf Q}$ is roughly $25\%$ higher than at
${\bf Q}'$. Despite the variation in $\Delta_{\pm,{\bf k}}$, there
is no anisotropy of the low-energy spin-wave velocity. The difference between
the results
for $\theta=0$ and $\theta=\pi/8$ at higher energies is due to the lower edge of
the continuum in the latter case.
Whereas the spin waves close to ${\bf Q}$ have a very similar dispersion
compared to the insulating model [\fig{EI1sw}(b)], the dispersion close to ${\bf
Q}'$ has two noticeable kinks at $\omega=0.65\,\Delta$ and
$\omega=1.25\,\Delta$. As shown in the inset, these kinks coincide with abrupt
changes in $\text{Im}\,{\cal D}({\bf q},\omega)$: $\text{Im}\,{\cal D}({\bf
  q},\omega)$ becomes finite at $\omega=0.65\,\Delta$, and starts to sharply
increase at $\omega=1.25\,\Delta$. The first feature corresponds to the
onset of Landau damping as the spin-wave branch enters the
continuum. The second feature is a result of the DOS, as discussed in the
following paragraph.

%a peculiarity of the structure of the continuum, which we discuss in the
%following paragraph.
%Note that the finite value of $\text{Im}\,{\cal D}({\bf q},\omega)$ for the
%spin waves close to ${\bf Q}$ and for $\omega<2\Delta$
%is an artifact of the finite width $\delta=0.1\,$meV.

\begin{figure}
  \includegraphics[width=0.9\columnwidth,clip]{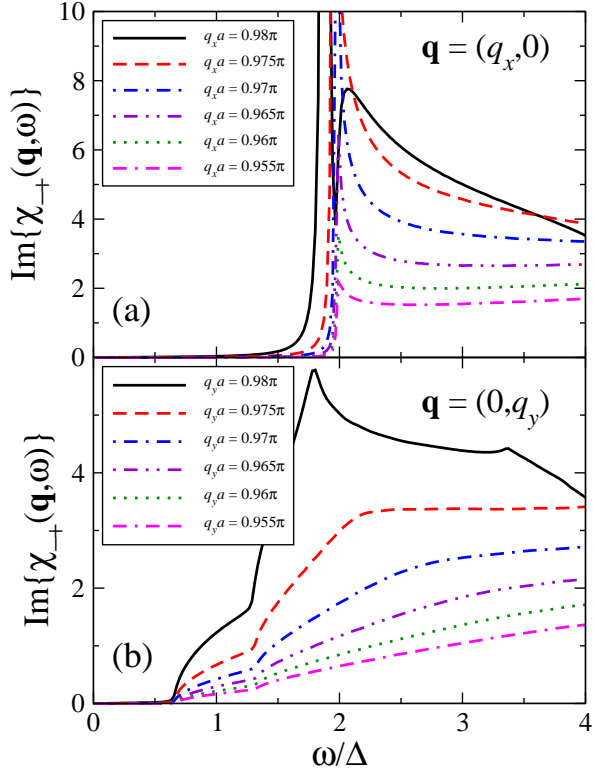}
  \caption{\label{slice}(Color online) Imaginary part of the transverse
    susceptibility as a function of $\omega$ at various values of ${\bf q}$
    near (a) ${\bf Q}$ and (b) ${\bf Q}'$ in the metallic excitonic
    model. We have calculated
  the mean-field susceptibilities using a $30000\times30000$ ${\bf k}$-point
  mesh and a width $\delta=0.2\,$meV.}
\end{figure}

Examining the lower edge of the continuum in both panels
of~\fig{interbchi6_1vs3}, we see that whereas the continuum disappears sharply
at $\omega=2\Delta$ near ${\bf Q}$, it appears to vanish
more smoothly near ${\bf Q}'$. In the latter case there are
two distinct thresholds, which are particularly visible around
$q_y=0.99\,\pi/a$. To examine this more closely, we plot
$\text{Im}\,\chi_{-+}({\bf q},\omega)$ as a function of $\omega$ for fixed ${\bf
  q}$ near ${\bf Q}$ and ${\bf Q}'$ in~\fig{slice}(a) and (b),
respectively. In the former case, we see the step-like
start of the continuum at $\omega=2\Delta$.
% which for all values of
%${\bf q}=(q_x,0)$ shown here is the minimum energy required to excite a
%particle from the states near the gapped Fermi pockets across the SDW
%gap. 
The peak at this energy is due both to the remnant of the spin-wave branch
(at least for $q_x=0.98\,\pi/a$) and to the enhancement of the DOS at the edge
of the SDW gap. The finite value of $\text{Im}\,\chi_{-+}({\bf q},\omega)$ for
$\omega<2\Delta$ is an artifact of the
finite width $\delta$. The susceptibility near ${\bf Q}'$ is
qualitatively different: the lower threshold of the continuum is at
$\omega_{1}=0.65\,\Delta$, and immediately above this the susceptibility
increases continuously as $\sqrt{\omega-\omega_1}$. At
$\omega_2=1.3\,\Delta$, the susceptibilility abruptly starts to increase more
steeply. The locations of these two thresholds correspond to the
kinks in the spin-wave dispersion. The rapid increase of
$\text{Im}\,\chi_{-+}({\bf q},\omega)$ above $\omega_2$ accounts for the strong
increase in the damping (see inset of~\fig{interbchi6_sw}).

As for the interband susceptibility near ${\bf Q}$, the origin of the lower
threshold is the variation of $\Delta_{\pm,{\bf k}}$. The difference is that
here the threshold originates from the minimum energy required for a
single-particle excitation between the states near the gapped hole pocket
and the states near the ungapped electron pocket,
$\omega_{1}=\min_{{\bf k}}(\Delta_{+,{\bf k}},\Delta_{-,{\bf
    k}})$. From~\fig{deltapm}(c) and (d) we deduce
$\omega_{1}\approx 0.6\,\Delta$, closely matching the lower threshold
in~\fig{slice}(b). The strong increase in $\text{Im}\,\chi_{-+}({\bf q},\omega)$
above $\omega_2$ is due to the peaks in the DOS located at
$\pm\omega_{2}$ on either side of the Fermi energy, shown
in~\fig{deltapm}(b): because of this DOS enhancement, the ``density of
excitations'' between states close to the gapped and ungapped Fermi pockets
is increased above $\omega_2$.

\section{Experimental situation}\label{sec:experiments}

This work ultimately aims to shed light upon the nature of the
antiferromagnetism in the iron pnictides, in particular
the extent to which it is
itinerant or localized in character. There are several
published results of inelastic neutron scattering examining the
spin excitations in the antiferromagnetic state of
CaFe$_2$As$_2$,\cite{McQueeney2008,Diallo2009,Zhao2009} 
SrFe$_2$As$_2$,\cite{Zhao2008} and
BaFe$_2$As$_2$.\cite{Ewings2008,Matan2009} 
%In particular, the magnetic order in all
%these materials is three-dimensional, with antiferromagnetic ordering of the
%spins at the iron sites along the $c$-axis. 
In these experiments, only transverse excitations contribute to the
neutron-scattering cross section, allowing us to write it as
\beq
\frac{d^2\sigma}{d\Omega{dE}} \propto |F({\bf q})|^2\, [n_B(\omega) +
  1]\,\text{Im}\,\chi_{-+}({\bf q},\omega),
\eeq
where $F({\bf q})$ is a form factor and $n_{B}(\omega)$ is the Bose-Einstein
distribution function. A direct, quantitative comparison between theory
and experiment would require a more realistic model for the low-energy band
structure than the one we are using. We nevertheless make
several general
remarks relating what we have learnt about the spin excitations in the
excitonic SDW state to the experimental results.

% interpretation of the neutron-scattering data: local moment or itinerant?
% where does the edge of the continuum begin? What is the value of Delta? Note
% that our work does not use relaistic band structure parameters. Perhaps
% better agreement can be found in this case...

We first review the experimental situation. Despite considerable variation in
the N\'eel temperature within the
$A$Fe$_2$As$_2$ ($A=$Ca, Sr, Ba) family, the static magnetic properties of
these compounds are rather similar. In particular, antiferromagnetism only
occurs in the presence of an orthorhombic distortion, which fixes the
ordering vector $\mathbf{Q}$. Experiments on the low-energy spin dynamics
are also in broad agreement: there is a strongly dispering spin wave close to
$\mathbf{Q}$,\cite{McQueeney2008,Diallo2009,Zhao2009,Zhao2008,Ewings2008,%
Matan2009} the spin-wave velocity is
anisotropic,\cite{McQueeney2008,Diallo2009,Zhao2009,Zhao2008,Matan2009} and
the spin-wave dispersion has a gap of energy
$6$--$10\,$meV.\cite{McQueeney2008,Diallo2009,Zhao2009,Zhao2008,Ewings2008,%
Matan2009}
At present, however, there is considerable disagreement over the high-energy
excitations. For CaFe$_2$As$_2$, it was reported\cite{Diallo2009} that the spin
wave is strongly damped at energies above
$100\,$meV, suggesting the presence of a particle-hole continuum. On the other
hand, although Zhao \emph{et al.}\cite{Zhao2009}\ found
similar spin-wave
velocities, they did not observe any significant jump in the damping of the
spin wave below $200\,$meV, which would indicate the intersection of the spin
wave dispersion with the continuum. The results for BaFe$_2$As$_2$ show greater
inconsistency, with reports\cite{Ewings2008} of strong spin excitations possibly
up to $170\,$meV in stark disagreement with claims of
spin-wave damping by continuum excitations at energies
as low as $24\,$meV.\cite{Matan2009}

The results of Refs.~\onlinecite{Matan2009} and \onlinecite{Diallo2009} are
most consistent with itinerant antiferromagnetism, as the existence of a
particle-hole continuum is a key feature of this scenario. Interpreting the  
latter experiment\cite{Diallo2009} in terms of the excitonic model, we deduce
a SDW gap of $\Delta\approx50\,$meV. This is nearly twice the estimate
$\Delta\approx30\,$meV of the $T=0$ gap based on ARPES for
SrFe$_2$As$_2$.\cite{Hsieh2008} Although a SDW gap of only $12\,$meV for
BaFe$_2$As$_2$, which we could infer from \Ref{Matan2009}, seems low, we have
seen above that spin-wave damping sets in at energies much smaller than
$2\Delta$, depending upon the details of the reconstructed band structure.  
In order to fit the results of~\Ref{Zhao2009} into the excitonic picture,
however, we require a SDW gap of at least $100\,$meV implying a rather high
value of the ratio $\Delta/k_{B}T_\text{SDW}\gtrsim7$. These results instead
support a local-moment
picture.\cite{Yildirim2008,Si2008,Uhrig2009,Krueger2009}  
The absence of the continuum is nevertheless surprising since
ARPES shows clear evidence for quasiparticle bands at low
energies, which suggests a possible resolution:~\cite{Hsieh2008,Lu2009} the
imperfect nesting of the elliptical electron pockets with the circular hole
pocket is expected to yield incompletely-gapped Fermi surfaces in the SDW
state, which implies that continuum excitations are present down to zero
energy. As such, the spin waves would be damped at all energies, and the jump
in the damping characteristic of the entry into the continuum is absent. Such
an explanation is of course at odds with~\Ref{Diallo2009}, indicating the need
for further work to clarify the experimental situation.

%Assuming a linear scaling of the zero temperature SDW gap with
%$T_{\text{SDW}}$,

% anisotropy of spin-wave dispersion: very hard to understand within our
% model. Possible effect due to orbital physics?

%The spin-wave velocities will be largely fixed by the details of the band
%structure, and more realistic calculations are needed to compare with the
%measured values. The anisotropy of the spin-wave velocity along
%the $c$-axis is not unreasonable in these layered quasi-2D
%compounds, and is not obviously inconsistent with the excitonic scenario.

The reported $40$\% anisotropy of the spin-wave velocity within the
$ab$ plane\cite{Zhao2009,Diallo2009} is quite remarkable.
Although this effect is absent from
our results due to the tetragonal symmetry of the Fermi pockets, it
nevertheless seems rather too large to be accounted for by the expected
elliptical shape of the electron-like Fermi pockets in the
pnictides.\cite{Zhao2009} Experimental results also do not show a second
spin-wave branch at ${\bf Q}'$, as found here for the metallic SDW model.
Both observations are likely due to the orthorhombic distortion in the SDW
phase, which lifts
the degeneracy of the $(\pi,0)$ and $(0,\pi)$ SDW,\cite{Brydon2009} 
and do not imply a failure of the excitonic scenario.

% spin gap in dispersion - coupling of magnetism to orthorhombic distortion of
% crystal, similarity to ordered phases of Managanese alloys

We finally remark upon the gap in the spin-wave dispersion in the
pnictides. Due to the absence of
of magnetic anisotropy is our model, we always find
Goldstone modes in the SDW phase. As demonstrated in Fishman and Liu's study
of manganese alloys,\cite{FishmanLiuMn} a gap is
possible in an excitonic SDW state in the presence of magnetoelastic coupling.
The magnetoelastic coupling in the pnictides is indeed strong, as evidenced by
the role of the orthorhombic distortion in fixing
the polarization and the ordering vector of the
SDW,\cite{1111coupling,122coupling,Matan2009} suggesting that it might be
responsible for the spin-wave gap.

% basic conclusion: not inconsistent with current results, but obviously more
% detailed analysis is required to be able to perform a quantitative
% comparison. 

In summary, the neutron-scattering data for the antiferromagnetic state in
the pnictides are currently unable to decide upon the origin and character of
the magnetism. We have shown that the excitonic SDW scenario gives spin-wave
excitations in qualitative agreement with experiments. An
obvious direction of future work is therefore to examine the spin excitations
based on more realistic band structures. Considering the imperfect
  nesting of the electron and hole pockets in the pnictides, it will be
  particularly interesting to address the possibility of incommensurate
  SDW order.~\cite{Cvetkovic2009}  The
effects of the interactions not directly contributing to the SDW instability
should be included. Comparison of our results with those obtained within
a model
explicitly accounting for the orbital character of the bands is also
important. Furthermore, the orthorhombic distortion and a
magnetoelastic coupling should be implemented for greater realism.
Although the spin excitations in more sophisticated models
will differ in their details from those presented here, we nevertheless think
that our
results will remain qualitatively correct and will thus be valuable in
interpreting 
future experiments.

%believe that the insight gained from our study will be valuable in
%understanding these results.

\section{Summary}\label{sec:summary}

We have presented an analysis of the zero-temperature transverse spin
excitations in the
excitonic SDW state of two-band, 2D models with nested electron-like and
hole-like Fermi pockets. Using the RPA, we have
derived the Dyson equation for the spin susceptibility and have shown that the
total spin susceptibility can be divided into contributions from interband and
intraband excitations. We have solved the Dyson equation in the 
special case when only the interactions responsible for the SDW are
non-zero. While the interband excitations are then directly enhanced by the
interactions, the intraband excitations are still indirectly enhanced due to the
mixing of the electron-like and hole-like states in the SDW phase.
The susceptibility exhibits collective spin-wave branches close to the SDW
ordering vector $\mathbf{Q}$ and also, with much smaller weight, close to
$\mathbf{q}=0$, as well as a continuum of single-particle excitations at
energies above a threshold of the order of the SDW gap.

Depending upon the non-interacting band structure, the opening of the
excitonic gap can result in qualitatively different SDW states. This has been
illustrated by considering two models, one which becomes insulating in the SDW
state and another which remains metallic due to the presence of an ungapped
portion of the Fermi surface. For comparison, we have also performed the
corresponding calculations for a 2D Hubbard model with the same mean-field SDW
gap. Differences in the spin excitations between the insulating and metallic
models occur only at low energies and mainly close to the nesting vector
$\mathbf{Q}'$ between the (gapped) hole pocket and the ungapped electron 
pocket, which is essentially unaffected by the SDW formation. We have also
discussed data from neutron-scattering experiments in light of our
results. We
conclude that the available data do not yet allow us to distinguish between an
excitonic SDW and a local-moment scenario for the antiferromagnetic order in
the pnictides.

% Future work: mention additional magnetic ordering behavior. Perhaps might
% explain the (pi/2,pi/2) ordering in FeTe or whatever.


\begin{thebibliography}{99}

\bibitem{Kamihara2008}Y. Kamihara, T. Watanabe, M. Hirano, and H. Hosono,
  J. Am.\ Chem.\ Soc.\ {\bf 130}, 3296 (2008).

\bibitem{Rotter2008}M. Rotter, M. Tegel, and D. Johrendt,
  Phys.\ Rev.\ Lett.\ {\bf 101}, 107006 (2008).

\bibitem{Ren2008}Z.-A. Ren, W. Lu, J. Yang, W. Yi, X.-L. Shen, Z.-C. Li,
  G.-C. Che, X.-L. Dong, L.-L. Sun, F. Zhou, and Z.-X. Zhao, Chin.\ Phys.\
  Lett.\ \textbf{25}, 2215 (2008).

\bibitem{Lynn2009}J. W. Lynn and P. Dai, Physica C \textbf{469}, 469 (2009).

\bibitem{delaCruz2008}C. de la Cruz, Q. Huang, J. W. Lynn, J. Li,
  W. Ratcliff II, J. L. Zarestky, H. A. Mook, G. F. Chen, J. L. Luo,
  N. L. Wang, P. Dai, Nature {\bf 453}, 899 (2008).

\bibitem{1111coupling}J. Zhao, Q. Huang, C. de la Cruz, S. Li,
  J. W. Lynn, Y. Chen, M. A. Green, G. F. Chen, G. Li, Z. Li, J. L. Luo,
  N. L. Wang, P. Dai, Nature Mater.\ {\bf{7}}, 953
  (2008); J. Zhao, Q. Huang, C. de la Cruz, J. W. Lynn, M. D. Lumsden,
  Z. A. Ren, J. Yang, X. Shen, X. Dong, Z. Zhao, P. Dai, Phys.\ Rev.\ B
  {\bf{78}}, 132504 (2008). 

\bibitem{122coupling}Q. Huang, Y. Qiu, W. Bao, J. W. Lynn, M. A. Green,
  Y. Chen, T. Wu, G. Wu, and X. H. Chen, Phys.\ Rev.\ Lett.\ {\bf 101},
  257003 (2008); A. Jesche, N. Caroca-Canales, H. Rosner, H. Borrmann,
  A. Ormeci, D. Kasinathan, K. Kaneko, H. H. Klauss, H. Luetkens, R. Khasanov,
  A. Amato, A. Hoser, C. Krellner, and C. Geibel, Phys.\ Rev.\ B {\bf{78}},
  180504(R) (2008).

\bibitem{Luetkens2009}H. Luetkens, H.-H. Klauss, M. Kraken, F. J. Litterst,
  T. Dellmann, R. Klingeler, C. Hess, R. Khasanov, A. Amato, C. Baines,
  M. Kosmala, O. J. Schumann, M. Braden, J. Hamann-Borrero, N. Leps,
  A. Kondrat, G. Behr, J. Werner, and B. B\"uchner, Nature Mater.\
  {\bf 8}, 305 (2009).

\bibitem{Mu2008}G. Mu, X.-Y. Zhu, L. Fang, L. Shan, C. Ren, and H.-H. Wen,
  Chin.\ Phys.\ Lett.\ \textbf{25}, 2221 (2008).

\bibitem{Shan2008}L. Shan, Y. Wang, X. Zhu, G. Mu, L. Fang, C. Ren, and H.-H.
  Wen, EPL \textbf{83}, 57004 (2008).

\bibitem{Si2008}Q. Si and E. Abrahams, Phys.\ Rev.\ Lett.\ {\bf 101}, 076401
  (2008).% theory, predict d_xy superconducting order

\bibitem{Kuroki2008}K. Kuroki, S. Onari, R. Arita, H. Usui, Y. Tanaka,
  H. Kontani, and H. Aoki, Phys.\ Rev.\ Lett.\ {\bf 101},
  087004 (2008).% theory, OK

%\bibitem{Korshunov2008}M. M. Korshunov and I. Eremin, EPL {\bf 83}, 67003
%  (2008);

\bibitem{Korshunov2008}M. M. Korshunov and I. Eremin, Phys.\ Rev.\ B {\bf 78},
  140509(R) (2008).% theory, OK

\bibitem{Lee2006}P. A. Lee, N. Nagaosa, and X.-G. Wen, Rev.\ Mod. Phys.\ {\bf
  78}, 17 (2006).

\bibitem{Singh2008}D. J. Singh and M.-H. Du, Phys.\ Rev.\ Lett.\ {\bf 100},
  237003 (2008);
  I. I. Mazin, D. J. Singh, M. D. Johannes, and M. H. Du,
  \textit{ibid.}\ {\bf 101}, 057003 (2008).
% mistake in Singh and Du reference in earlier version

%\bibitem{Khasanov2009}R. Khasanov, D. V. Evtushinsky, A. Amato, H.-H. Klauss,
%  H. Luetkens, C. Niedermayer, B. B\"{u}chner, G. L. Sun, C. T. Lin,
%  J. T. Park, D. S. Inosov, and V. Hinkov, Phys.\ Rev.\ Lett.\ {\bf 102},
% 187005 (2009).

%\bibitem{Daghero2009}D. Daghero, M. Tortello, R. S. Gonnelli, V. A. Stepanov,
%  N. D. Zhigadlo, and J. Karpinski, Phys.\ Rev.\ B {\bf 80}, 060502(R) (2009).

%\bibitem{Evtushinsky2009}D. V. Evtushinsky, D. S. Inosov, V. B. Zabolotnyy,
%  M. S. Viazovska, R. Khasanov, A. Amato, H.-H. Klauss, H. Luetkens,
%  C. Niedermayer, G. L. Sun, V. Hinkov, C. T. Lin, A. Varykhalov, A.Koitzsch,
%  M. Knupfer, B. B\"{u}chner, A. A. Kordyuk, and S. V. Borisenko, New
%  J. Phys.\ {\bf 11}, 055069 (2009).

\bibitem{McGuire2008}M. A. McGuire, A. D. Christianson, A. S. Sefat,
  B. C. Sales, M. D. Lumsden, R. Jin, E. A. Payzant, D. Mandrus, Y. Luan,
  V. Keppens, V. Varadarajan, J. W. Brill, R. P. Hermann, M. T. Sougrati,
  F. Grandjean, and G. J. Long, Phys.\ Rev.\ B {\bf 78}, 094517 (2008);
  M. A. McGuire, R. P. Hermann, A. S. Sefat, B. C. Sales, R. Jin, D. Mandrus,
  F. Grandjean, and G. J. Long, New J. Phys.\ {\bf 11}, 025011 (2009).% both OK

\bibitem{Liu2008}R. H. Liu, G. Wu, T. Wu, D. F. Fang, H. Chen, S. Y. Li,
  K. Liu, Y. L. Xie, X. F. Wang, R. L. Yang, L. Ding, C. He, D. L. Feng, and
  X. H. Chen, Phys.\ Rev.\ Lett.\ {\bf 101}, 087001 (2008).% OK

\bibitem{Dong2008}J. K. Dong, L. Ding, H. Wang, X. F. Wang, T. Wu, G. Wu,
X. H. Chen, and S. Y. Li, New J. Phys.\ \textbf{10}, 123031 (2008).% added (122)

\bibitem{magneto}S. E. Sebastian, J. Gillett, N. Harrison, P. H. C. Lau,
  C. H. Mielke, and G. G. Lonzarich, J. Phys:
  Condens.\ Matter {\bf{20}}, 422203 (2008); J. G. Analytis, R.
  D. McDonald, J.-H. Chu, S. C. Riggs, A. F. Bangura, C. Kucharczyk,
  M. Johannes, and I. R. Fisher, Phys.\ Rev.\ B {\bf 80}, 064507 (2009).

\bibitem{Hsieh2008}D. Hsieh, Y. Xia, L. Wray, D. Qian, K. Gomes, A. Yazdani,
  G. F. Chen, J. L. Luo, N. L. Wang, and M. Z. Hasan, arXiv:0812.2289v1
  (unpublished).

\bibitem{Boeri}L. Boeri, O. V. Dolgov, and A. A. Golubov,
  Phys.\ Rev.\ Lett.\ {\bf 101}, 026403 (2008); Physica C {\bf 469}, 628
  (2009).

\bibitem{Fedders1966}P. A. Fedders and P. C. Martin, Phys.\ Rev.\ {\bf 143}, 245
  (1966).

\bibitem{Liu1970}S. H. Liu, Phys.\ Rev.\ B {\bf 2}, 2664 (1970).

\bibitem{Rice1970}T. M. Rice, Phys.\ Rev.\ B {\bf 2}, 3619 (1970).

\bibitem{Kulikov1984}N. I. Kulikov and V. V. Tugushev, Sov.\ Phys.\ Usp.\ {\bf
  27}, 954 (1984).

\bibitem{FawcettCr}E. Fawcett, Rev.\ Mod. Phys.\ {\bf 60}, 209 (1988);
  E. Fawcett, H. L. Alberts, V. Y. Galkin, D. R. Noakes, and J. V. Yakhmi,
  Rev.\ Mod.\ Phys.\ {\bf 66}, 25 (1994).

%\bibitem{FishmanLiuCr}R. S. Fishman and S. H. Liu, Phys.\ Rev.\ B {\bf 50},
%  R4240 (1994); {\bf 54}, 7233 (1996).

\bibitem{FishmanLiuMn}R. S. Fishman and S. H. Liu, Phys.\ Rev.\ B {\bf 58},
  R5912 (1998); {\bf 59}, 8672 (1999); {\bf 59}, 8681 (1999).

\bibitem{Excitonic}L. V. Keldysh and Y. V. Kopaev, Sov.\ Phys.\ Solid State {\bf
  6}, 2219 (1965); J. des Cloizeaux, J. Phys.\ Chem.\ Solids {\bf 26}, 259
  (1965); D. J\'erome, T. M. Rice, and W. Kohn, Phys.\ Rev.\ {\bf 158}, 462
  (1967).

\bibitem{Volkov1976}B. A. Volkov, Y. V. Kopaev, and A. I. Rusinov, Sov.\
  Phys.\ JETP {\bf 41}, 952 (1976).

\bibitem{Buker1981}D. W. Buker, Phys.\ Rev.\ B {\bf{24}}, 5713 (1981).

\bibitem{Chubukov2008}A. V. Chubukov, D. Efremov, and I. Eremin, Phys.\ Rev.\ B
  {\bf{78}}, 134512 (2008).

\bibitem{Han2008}Q. Han, Y. Chen, and Z. D. Wang, EPL {\bf{82}},
  37007 (2008).

\bibitem{Mizokawa2008}T. Mizokawa, T. Sudayama, and Y. Wakisaka,
  J. Phys.\ Soc.\ Jpn.\ Suppl.\ C {\bf 77}, 158 (2008).

\bibitem{Cvetkovic2009}V. Cvetkovic and Z. Tesanovic Europhys. Lett. {\bf 85},
  37002, (2009); Phys. Rev. B {\bf 80}, 024512 (2009). 

\bibitem{Vorontsov2009}A. B. Vorontsov, M. G. Vavilov, and A. V. Chubukov,
  Phys.\ Rev.\ B {\bf 79}, 060508(R) (2009).

\bibitem{Brydon2009}P. M. R. Brydon and C. Timm, Phys.\ Rev.\ B {\bf 79},
  180504(R) (2009).


\bibitem{Raghu2008}S. Raghu, Z.-L. Qi, C.-X. Liu, D. J. Scalapino, and
  S.-C. Zhang, Phys.\ Rev.\ B {\bf 77}, 220503(R) (2008).

\bibitem{Lorenzana2008}J. Lorenzana, G. Seibold, C. Ortix, and M. Grilli,
  Phys.\ Rev.\ Lett.\ {\bf 101}, 186402 (2008).

\bibitem{Ran2009}Y. Ran, F. Wang, H. Zhai, A. Vishwanath, and D.-H. Lee,
  Phys.\ Rev.\ B {\bf 79}, 014505 (2009).

\bibitem{Yu2009}R. Yu, K. T. Trinh, A. Moreo, M. Daghofer, J. A. Riera,
  S. Haas, and E. Dagotto, Phys.\ Rev.\ B {\bf 79}, 104510 (2009).

\bibitem{Yildirim2008}T. Yildirim, Phys.\ Rev.\ Lett.\ {\bf 101}, 057010 (2008).

\bibitem{Uhrig2009}G. S. Uhrig, M. Holt, J. Oitmaa, O. Sushkov, and
  R. P. P. Singh, Phys.\ Rev.\ B {\bf 79}, 092416 (2009).

\bibitem{Krueger2009}F. Kr\"{u}ger, S. Kumar, J. Zaanen, and J. van den Brink,
  Phys.\ Rev.\ B {\bf 79}, 054504 (2009).

\bibitem{Kroll2008}T. Kroll, S. Bonhommeau, T. Kachel, H. A. D\"urr, J. Werner,
  G. Behr, A. Koitzsch, R. H\"ubel, S. Leger, R. Sch\"onfelder, A. Ariffin,
  R. Manzke, F. M. F. de Groot, J. Fink, H. Eschrig, B. B\"uchner, and
  M. Knupfer, Phys.\ Rev.\ B {\bf 78}, 220502 (2008).

\bibitem{Yang2009}W. L. Yang, P. O. Velasco, J. D. Denlinger, A. P. Sorini,
  C-C. Chen, B. Moritz, W.-S. Lee, F. Vernay, B. Delley, J.-H. Chu,
  J. G. Analytis, I. R. Fisher, Z. A. Ren, J. Yang, W. Lu, Z. X. Zhao, J. van
  den Brink, Z. Hussain, Z.-X. Shen, and T. P. Devereaux, Phys.\ Rev.\ B {\bf
  80}, 014508 (2009).

\bibitem{Ewings2008}R. A. Ewings, T. G. Perring, R. I. Bewley, T. Guidi,
  M. J. Pitcher, D. R. Parker, S. J. Clarke, and A. T. Boothroyd, Phys.\ Rev.\ B
  {\bf 78}, 220501(R) (2008).

\bibitem{Zhao2009}J. Zhao, D. T. Adroja, D.-X. Yao, R. Bewley, S. Li,
  X. F. Wang, G. Wu, X. H. Chen, J. Hu, and P. Dai, Nature Phys.\ {\bf 5},
  555 (2009).

\bibitem{Schrieffer1989}J. R. Schrieffer, X. G. Wen, and S. C. Zhang,
  Phys.\ Rev.\ B {\bf 39}, 11663 (1989).

\bibitem{Singh1990}A. Singh and Z. Te\v{s}anovi\'{c}, Phys.\ Rev.\ B {\bf 41},
  614 (1990).

\bibitem{Chubukov1992}A. V. Chubukov and D. M. Frenkel, Phys.\ Rev.\ B {\bf 46},
  11884 (1992).

\bibitem{Anderson1952}P. W. Anderson, Phys.\ Rev.\ {\bf 86}, 694 (1952).
% OK, this is the proper reference and is cited by others in the same context.
% The paper is not quite easy to understand, though.

\bibitem{Hasegawa1978}H. Hasegawa, J. Low Temp.\ Phys.\ {\bf 31}, 475 (1978).

%\bibitem{TiSe2}H. Cercellier, C. Monney, F. Clerc, C. Battaglia, L. Despont,
%  M. G. Garnier, H. Beck, P. Aebi, L. Patthey, H. Berger, and L. Forr\'{o},
%  Phys.\ Rev.\ Lett.\ {\bf 99}, 146403 (2007); C. Monney, H. Cercellier,
%  F. Clerc, C. Battaglia, E. F. Schwier, C. Didiot, M. G. Garnier, H. Beck,
%  P. Aebi, H. Berger, L. Forr\'{o}, and L. Patthey, Phys.\ Rev.\ B {\bf 79},
%  045116 (2009).

%\bibitem{Ta2NiSe5}Y. Wakisaka, T. Sudayama, K. Takubo, T. Mizokawa, M. Arita,
%  H. Namatama, M. Taniguchi, N. Katayama, M. Nohara, and H. Takagi,
%  Phys.\ Rev.\ Lett.\ {\bf 103}, 026402 (2009).

\bibitem{Batista2002}C. D. Batista, Phys. Rev. Lett. {\bf 89}, 166403 (2002).

\bibitem{McQueeney2008}R. J. McQueeney, S. O. Diallo, V. P. Antropov,
  G. Samolyuk, C. Broholm, N. Ni, S. Nandi, M. Yethiraj, J. L. Zarestky,
  J. J. Pulikkotil, A. Kreyssig, M. D. Lumsden, B. N. Harmon, P. C. Canfield,
  and A. I. Goldman, Phys.\ Rev.\ Lett.\ {\bf 101}, 227205
(2008).

\bibitem{Diallo2009}S. O. Diallo, V. P. Antropov, T. G. Perring, C. Broholm,
  J. J. Pulikkotil, N. Ni, S. L. Bud'ko, P. C. Canfield, A. Kreyssig,
  A. I. Goldman, and R. J. McQueeney, Phys.\ Rev.\ Lett.\ {\bf
  102}, 187206 (2009).

\bibitem{Zhao2008}J. Zhao, D.-X. Yao, S. Li, T. Hong, Y. Chen, S. Chang,
  W. Ratcliff II, J. W. Lynn, H. A. Mook, G. F. Chen, J. L. Luo, N. L. Wang,
  E. W. Carlson, J. Hu, and P. Dai, Phys.\ Rev.\ Lett.\ {\bf 101}, 167203
  (2008).

\bibitem{Matan2009}K. Matan, R. Morinaga, K. Iida, and T. J. Sato,
  Phys.\ Rev.\ B {\bf 79}, 054526 (2009).

\bibitem{Lu2009}D. H. Lu, M. Yi, S.-K. Mo, J. G. Analytis, J.-H. Chu,
  A. S. Erickson, D. J. Singh, Z. Hussain, T. H. Geballe, I. R. Fisher,
  and Z.-X. Shen, Physica C {\bf 469}, 452 (2009).



%\bibitem{Mazin2009}I. I. Mazin and J. Schmalian, arXiv:0901.4709
%  (unpublished).

\end{thebibliography}
\end{document}